\documentclass[aps,pre,superscriptaddress]{revtex4}
\usepackage[dvips]{graphics}
\usepackage{graphicx}
\usepackage{amsfonts}
\usepackage{amssymb}
\usepackage{amsmath}

\begin{document}

\title{PT-symmetry Management in Oligomer Systems}

\author{R.L. Horne}
\affiliation{Department of Mathematics, Morehouse College, Atlanta, GA 30314}
\email{rhorne@morehouse.edu}

\author{J. Cuevas}
\affiliation{Grupo de F\'{i}sica No Lineal. Universidad de Sevilla. Departamento de F\'{i}sica Aplicada I. Escuela Polit\'ecnica Superior. C/ Virgen de \'Africa, 7. 41011 Sevilla, Spain}
\email{jcuevas@us.es}

\author{P.G.\ Kevrekidis }
\affiliation{Department of Mathematics and Statistics, University of Massachusetts,
Amherst MA 01003-4515, USA}
\email{kevrekid@gmail.com}

\author{N. Whitaker}
\affiliation{Department of Mathematics and Statistics, University of Massachusetts,
Amherst MA 01003-4515, USA}
\email{whitaker@math.umass.edu}

\author{F.Kh. Abdullaev}
\affiliation{Instituto de F\'{\i}sica T{\'e}orica, Universidade Estadual Paulista, 01140-070, S\~ao Paulo, S\~ao Paulo, Brazil}

\author{D.J. Frantzeskakis}
\affiliation{Department of Physics, University of Athens, Panepistimiopolis, Zografos, GR-15784 Athens, Greece}

\begin{abstract}
We study the effects of management of the PT-symmetric
part of the potential within the setting of  Schr{\"o}dinger
dimer and trimer oligomer systems. This is done by rapidly modulating in time
the gain/loss profile. This gives rise to a number of interesting properties
of the system, which are explored at the level of an averaged equation
approach. Remarkably, this rapid modulation provides for a controllable
{\it expansion} of the region of exact PT-symmetry, depending on the
strength and frequency of the imposed modulation.
The resulting averaged models
are analyzed theoretically and their exact stationary
solutions are translated into time-periodic solutions through the averaging
reduction. These are, in turn, compared with the exact periodic solutions
of the full non-autonomous PT-symmetry managed problem and very good
agreement is found between the two.
\end{abstract}

\maketitle

\section{Introduction}

It has been about a decade and a half since the radical and
highly innovative proposal of C. Bender and his collaborators~\cite{R1}
regarding the potential physical relevance of Hamiltonians
respecting Parity (P) and time-reversal (T) symmetries. While
earlier work was focused on an implicit postulate of solely
self-adjoint Hamiltonian operators, this proposal suggested that
these fundamental symmetries may allow for a real operator spectrum
within a certain regime of parameters which is regarded as the
regime of exact PT-symmetry. On the other hand, beyond a critical
parametric strength, the relevant operators may acquire a spectrum
encompassing imaginary or even genuinely complex eigenvalues, in which
case, we are referring (at the linear level) to the regime of
broken PT-phase.

These notions were intensely studied at the quantum mechanical
level, chiefly as theoretical constructs. Yet, it was the fundamental
realization that optics can enable such ``open'' systems featuring
gain and loss, both at the theoretical~\cite{Muga,ziad,Ramezani,Kuleshov} and
even at the experimental~\cite{dncnat,salamo} level, that propelled this
activity into a significant array of new directions, including the
possibility of the interplay of nonlinearity with PT-symmetry.
In this optical context, the well-known connection of the Maxwell
equations with the Schr{\"o}dinger equation was utilized, and
Hamiltonians of the form  $H=-(1/2) \Delta + V(x)$ were considered
at the linear level with the PT-symmetry necessitating that the potential
satisfies the condition  $V(x)=V^{\star}(-x)$. Yet another physical
context where such systems have been experimentally ``engineered''
recently is that of electronic circuits; see the work
of~\cite{tsampikos_recent} and also the review of~\cite{tsampikos_review}.
In parallel to the recent experimental developments, numerous
theoretical groups have explored various features of both
linear PT-symmetric
potentials~\cite{kot1,sukh1,kot2,grae1,grae2,kot3,pgk,dmitriev1,dmitriev2,R30add1,R30add2,R30add3,R30add4,R30add5,R34,R44,R46,baras1,baras2,konorecent3,leykam,konous,djf,kondark,uwe,tyugin,pickton} and even of nonlinear ones such where a PT-symmetric type of gain/loss pattern appears in the nonlinear
term~\cite{miron,konorecent,konorecent2,meid}.

Our aim in the present work is to combine this highly active research theme
of PT-symmetry with another topic of considerable recent interest in the
physics of optical and also atomic systems, namely that of ``management'';
see, e.g., Refs.~\cite{boris_book,turitsyn_review} for recent reviews.
Originally,
the latter field had a significant impact at the level of providing for
robust soliton propagation in suitable regimes of the so-called dispersion/nonlinearity
management. More recently, as the above references indicate, the possibility
(in both nonlinear optics and atomic physics) of periodic --or other--
variation also of the nonlinearity has become a tool of significant value
and has enabled to overcome a number of limitations including e.g. the
potential of catastrophic collapse of bright solitary waves in higher dimensions.
In our PT-symmetric setting, to the best of our knowledge, such a temporal modulation
of (just the linear in our case~\cite{footnote})
gain and loss over time has not been proposed previously. Admittedly, in the optical
setting, and over the propagation distance, this type of variation
may be harder to achieve. Nevertheless, in an electronic setting
where the properties of gain can be temporally controlled by relevant
switching devices, such a realization may be deemed as more feasible. Our
argument herein is that it is also very worthwhile to consider this problem from
the point of view of its implications.

In particular, in what follows, we illustrate that in the case of a
rapid modulation (``strong'' management \cite{boris_book,turitsyn_review}),
it is possible to understand the non-autonomous
PT-symmetric system by considering its effective averaged form.
We showcase this type of averaging in the case of PT-symmetric
oligomers, previously explored in a number of works
(see e.g.~\cite{kot1,pgk,konorecent3,tyugin,pickton,meid}, among others).
We examine, more specifically, the case of dimers and trimers~\cite{kipnew}
which are the most tractable (also analytically) among the relevant
configurations. Our findings suggest
that there are interesting features that arise in
the averaged models which are, in turn, found to be confirmed
by the original non-autonomous ones.
For instance, in the case of the dimer, the averaged effective model
develops an effective linear coupling (and nonlinear self-interaction)
coefficient, which has a dramatic implication in controllably expanding
the region of the exact PT-symmetric phase, as a function of the
strength and frequency of the associated modulation.
Our analysis clearly illustrates how this is a direct
consequence of the averaging process, and the properties of the
periodic solutions are reconstructed on the basis
of the averaging and are favourably compared to the observations
of the time-periodic solutions of the original non-autonomous
system.

Our presentation is structured
as follows. In section II, we systematically develop the averaging
procedure both for the dimer and for the trimer; the generalization
to more sites will then be evident. In section III, we provide some
general insight on the existence and stability of solutions in these
effective averaged systems. In section IV, we corroborate these results
with numerical simulations of the full non-autonomous dimer/trimer systems,
finding very good agreement between the two. Finally, in section V,
we summarize our conclusions and comment on some interesting directions
for potential future work.

\section{The Averaged Equations for the PT-symmetric Dimer and Trimer models}

\subsection{DNLS PT-symmetric Dimer Model}

We start by considering
the DNLS dimer model with a rapidly-varying gain/loss term of the form:
\begin{eqnarray} \label{dnls1}
    i\frac{du}{dt} = \kappa v + |u|^{2}u + i\gamma_{0}u + \frac{i}{\epsilon}\gamma_{1}(t/\epsilon)u, \nonumber\\
        \nonumber \\
    i\frac{dv}{dt} = \kappa u + |v|^{2}v - i\gamma_{0}v - \frac{i}{\epsilon}\gamma_{1}(t/\epsilon)v,
 \end{eqnarray}
where $t$ is the evolution variable, $\epsilon$ is a small parameter, $\gamma_{0}$ represents the linear gain and loss  strength and $\gamma_{1}(t/\epsilon)$ is the rapidly-varying gain/loss profile that
will be central to our considerations herein.
We now apply a multiple scales analysis to Eq.~($\ref{dnls1}$) in
order to derive an averaged equation for our problem. First, we define the new
variables $\tau = t/\epsilon$ (fast scale) and $T = t$ (slow scale), and introduce the
the transformations
\begin{eqnarray} \label{dnls_var1}
     u(t) = U(T, \tau)\exp[\Gamma(\tau)], \qquad
     v(t) = V(T, \tau)\exp[-\Gamma(\tau)],
\end{eqnarray}
where $\Gamma(\tau)=\int_{0}^{\tau}~\gamma_{1}(\tau')d\tau'$. This way, we cast Eqs.~(\ref{dnls1}) into the following form:
\begin{eqnarray} \label{multiple_scales_eqn1}
    \frac{i}{\epsilon}\frac{\partial U}{\partial \tau} + i\frac{\partial U}{\partial T} = \kappa \exp[-2\Gamma(\tau)] V
     + \exp[2\Gamma(\tau)]|U|^{2}U + i\gamma_{0}U, \nonumber \\
      \nonumber \\
    \frac{i}{\epsilon}\frac{\partial V}{\partial \tau} + i\frac{\partial V}{\partial T} = \kappa \exp[2\Gamma(\tau)] U
     + \exp[-2\Gamma(\tau)]|V|^{2}V - i\gamma_{0}V.
 \end{eqnarray}
Next, expanding the unknown fields $U(T, \tau)$ and $V(T, \tau)$ in powers of $\epsilon$, i.e.,
\begin{eqnarray} \label{expansion1}
U(T, \tau) = \sum_{n=0}^{\infty}~\epsilon^{n}U_{n}(T, \tau), \qquad
V(T, \tau) = \sum_{n=0}^{\infty}~\epsilon^{n}V_{n}(T, \tau),
\end{eqnarray}
we derive from Eqs.~($\ref{multiple_scales_eqn1}$) the following results.

At the leading-order of approximation, i.e., at $\mathcal{O}(1/\epsilon)$, we obtain the equations
$i\partial U_0/\partial \tau=0$ and $i\partial V_0/\partial \tau=0$, which suggest that the fields $U_0$ and $V_0$
depend only on the slow time scale $T$, i.e.,
\begin{equation}
U_{0}(T, \tau) = \tilde{U}_{0}(T), \qquad V_{0}(T, \tau) = \tilde{V}_{0}(T).
\label{Oe-1}
\end{equation}
Additionally, at the order $\mathcal{O}(1)$, we
obtain the following set of equations:
\begin{eqnarray}
i\frac{\partial U_{1}}{\partial \tau} &+& i\frac{\partial U_{0}}{\partial T} = \kappa \exp[-2\Gamma(\tau)]V_{0}
+ \exp[2\Gamma(\tau)]|U_{0}|^{2}U_{0} + i\gamma_{0}U_{0},
\nonumber \\
i\frac{\partial V_{1}}{\partial \tau} &+& i\frac{\partial V_{0}}{\partial T} = \kappa \exp[2\Gamma(\tau)]U_{0}
+ \exp[-2\Gamma(\tau)]|V_{0}|^{2}V_{0} - i\gamma_{0}V_{0}.
\label{O1}
\end{eqnarray}
Next, using the definition of the average of some function $f(\tau)$ over a period $T$ as
$\langle f(\tau) \rangle \equiv (1/T)\int_{0}^{T}~f(\tau)d\tau$,
we average Eqs.~(\ref{O1})
over the period $T_0$ of
$\gamma_{1}(\tau)$, and obtain the equations:
\begin{eqnarray}
  \frac{i}{T_{0}}\int_{0}^{T_{0}}~\frac{\partial U_{1}}{\partial \tau} d\tau = \frac{1}{T_{0}}\int_{0}^{T_{0}}~d\tau \left[ -i\frac{\partial \tilde{U}_{0}}{\partial T}
     + \kappa \exp[-2\Gamma(\tau)] \tilde{V}_{0} + \exp[2\Gamma(\tau)] |\tilde{U}_{0}|^{2}\tilde{U}_{0} + i\gamma_{0}\tilde{U}_{0}\right] \nonumber \\
        \nonumber \\
  \frac{i}{T_{0}}\int_{0}^{T_{0}}~\frac{\partial V_{1}}{\partial \tau} d\tau = \frac{1}{T_{0}}\int_{0}^{T_{0}}~d\tau \left[ -i\frac{\partial \tilde{V}_{0}}{\partial T}
     + \kappa \exp[2\Gamma(\tau)] \tilde{U}_{0} + \exp[-2\Gamma(\tau)] |\tilde{V}_{0}|^{2}\tilde{V}_{0} - i\gamma_{0}\tilde{V}_{0}\right],
\end{eqnarray}
where we have also used the result in Eqs.~(\ref{Oe-1}).
The solvability condition for these equations is satisfied if
derive the following set of averaged equations for $\tilde{U}_{0}(T)$ and $\tilde{V}_{0}(T)$:
\begin{eqnarray}
\label{dimer_averaged_eqns1}
   i\frac{\partial U_{0}}{\partial T} = \kappa_{1} V_{0} + g_{1}|U_{0}|^{2}U_{0} + i\gamma_{0}U_{0}, \nonumber \\
      \nonumber \\
     i\frac{\partial V_{0}}{\partial T} = \kappa_{2} U_{0} + g_{2}|V_{0}|^{2}V_{0} - i\gamma_{0}V_{0},
\end{eqnarray}
where, for convenience, we have dropped the tildes; the (constant) coefficients of the above system
are given by:
\begin{eqnarray} \label{dimer_parameters1}
    \kappa_{1} \equiv \frac{\kappa}{T_{0}}\int_{0}^{T_{0}}\exp[-2\Gamma(\tau)] d\tau,
    ~~~~~ \kappa_{2} \equiv \frac{\kappa}{T_{0}}
       \int_{0}^{T_{0}}\exp[2\Gamma(\tau)] d\tau, \hspace*{0.5in} \nonumber \\
       \nonumber \\
    g_{1} \equiv \frac{1}{T_{0}}\int_{0}^{T_{0}}\exp[2\Gamma(\tau)] d\tau = \frac{\kappa_{2}}{\kappa}, ~~~~~  g_{2} \equiv \frac{1}{T_{0}}
        \int_{0}^{T_{0}}\exp[-2\Gamma(\tau)] d\tau = \frac{\kappa_{1}}{\kappa},
\end{eqnarray}
and it should be recalled that $\Gamma(\tau) = \int_{0}^{\tau}~\gamma_{1}(\tau')d\tau'$ for some choice of
$\gamma_{1}(\tau)$. For example,
the choice of
$\gamma_{1}(\tau) = \gamma_{1}\cos(\tau)$ allows one to express $\kappa_{1}, \kappa_{2}, g_{1} ~\mbox{and}~ g_{2}$ in terms
of modified Bessel functions. In this case, the period $T_{0} = 2\pi$ and $\gamma_{1}$
is a constant that controls the amplitude of the temporal modulation.

In the next section, we will follow
the method employed above
to derive an averaged set of equations for the DNLS PT-symmetric
trimer model.


\subsection{DNLS PT-symmetric trimer model}

We now consider
the following DNLS trimer model with a rapidly-varying gain/loss term:
\begin{eqnarray} \label{dnls_trimer1}
    i\frac{du}{dt} = -\kappa v - |u|^{2}u - i\gamma_{0}u - \frac{i}{\epsilon}\gamma_{1}(t/\epsilon)u, 
        \nonumber \\
    i\frac{dv}{dt} = -\kappa (u + w) - |v|^{2}v, \hspace*{0.5in} 
       \nonumber \\
    i\frac{dw}{dt} = -\kappa v - |w|^{2}w + i\gamma_{0}w + \frac{i}{\epsilon}\gamma_{1}(t/\epsilon)w,
 \end{eqnarray}
where we have used the same notation as in the case of the dimer.
We again assume that the unknown fields depend on the fast and slow scales
$\tau$ and $T$, and can be expressed as:
\begin{eqnarray} \label{dnls_trimer_var1}
     u(t) = U(T, \tau)\exp[\Gamma(\tau)],  ~~~~~v(t) = V(T, \tau),~~~~~w(t) = W(T, \tau)\exp[-\Gamma(\tau)],
 \end{eqnarray}
where $\Gamma(\tau) = -\int_{0}^{\tau}~\gamma_{1}(\tau')d\tau'$, and
$U(T, \tau)$, $V(T, \tau)$ and $W(T, \tau)$ obey the following system:

 \begin{eqnarray} \label{multiple_scales_eqn2}
    \frac{i}{\epsilon}\frac{\partial U}{\partial \tau} + i\frac{\partial U}{\partial T} = -\kappa \exp[-\Gamma(\tau)] V
     - \exp[2\Gamma(\tau)]|U|^{2}U - i\gamma_{0}U, 
      \nonumber \\
    \frac{i}{\epsilon}\frac{\partial V}{\partial \tau} + i\frac{\partial V}{\partial T} = -\kappa \exp[\Gamma(\tau)] U - \kappa \exp[-\Gamma(\tau)] W
               - |V|^{2}V, 
        \nonumber \\
    \frac{i}{\epsilon}\frac{\partial W}{\partial \tau} + i\frac{\partial W}{\partial T} = -\kappa \exp[\Gamma(\tau)] V
     - \exp[-2\Gamma(\tau)]|W|^{2}W + i\gamma_{0}W.
 \end{eqnarray}
Next,
expanding, as before, $U(T, \tau)$, $V(T, \tau)$ and $W(T, \tau)$ in powers of $\epsilon$, namely,
 \begin{eqnarray} \label{expansion2}
     U(T, \tau) = \sum_{n=0}^{\infty}~\epsilon^{n}U_{n}(T, \tau), \qquad
     V(T, \tau) = \sum_{n=0}^{\infty}~\epsilon^{n}V_{n}(T, \tau), \qquad
     W(T, \tau) = \sum_{n=0}^{\infty}~\epsilon^{n}W_{n}(T, \tau),
 \end{eqnarray}
we obtain from Eqs.~($\ref{multiple_scales_eqn2}$) the following results.

First, at the order $\mathcal{O}(1/\epsilon)$, we obtain the equations
$i\partial U_0/\partial \tau=0$, $i\partial V_0/\partial \tau=0$, and
$i\partial W_0/\partial \tau=0$, which show that the fields $U_0$, $V_0$ and $W_0$
depend only on the slow time scale $T$, i.e.,
\begin{equation}
U_{0}(T, \tau) = \tilde{U}_{0}(T), \qquad V_{0}(T, \tau) = \tilde{V}_{0}(T), \qquad
W_{0}(T, \tau) = \tilde{W}_{0}(T).
\label{Oe-1t}
\end{equation}
Next, at the order $\mathcal{O}(1)$, we obtain the system:
\begin{eqnarray}
i\frac{\partial U_{1}}{\partial \tau} + i\frac{\partial U_{0}}{\partial T} = -\kappa \exp[-\Gamma(\tau)]V_{0}
                  - \exp[2\Gamma(\tau)]|U_{0}|^{2}U_{0} - i\gamma_{0}U_{0}, 
                 \nonumber \\
i\frac{\partial V_{1}}{\partial \tau} + i\frac{\partial V_{0}}{\partial T} = -\kappa \exp[\Gamma(\tau)] U_{0}
                    - \kappa \exp[-\Gamma(\tau)] W_{0} - |V_{0}|^{2}V_{0}, 
                 \nonumber \\
i\frac{\partial W_{1}}{\partial \tau} + i\frac{\partial W_{0}}{\partial T} = -\kappa \exp[\Gamma(\tau)]V_{0}
                  - \exp[-2\Gamma(\tau)]|W_{0}|^{2}W_{0} + i\gamma_{0}W_{0}.
              \end{eqnarray}
Similarly to the case for the PT-symmetric dimer model,
we average the above system over the period $T_0$ of
$\gamma_{1}(\tau)$.
Then, employing the solvability conditions for the resulting system, i.e.,
$U_{1}(T, \tau)$, $V_{1}(T, \tau)$ and $W_{1}(T, \tau)$ are periodic in $\tau$
with period $T_{0}$, we
obtain the following set of averaged equations for $\tilde{U}_{0}$, $\tilde{V}_{0}$ and $\tilde{W}_{0}$:
\begin{eqnarray} \label{trimer_averaged_eqns1}
   i\frac{\partial U_{0}}{\partial T} = -\tilde{\kappa}_{1} V_{0} - g_{1}|U_{0}|^{2}U_{0} - i\gamma_{0}U_{0}, 
      \nonumber \\
   i\frac{\partial V_{0}}{\partial T} = -\tilde{\kappa}_{2} U_{0} - \tilde{\kappa}_{1} W_{0} - |V_{0}|^{2}V_{0}, 
      \nonumber \\
   i\frac{\partial W_{0}}{\partial T} = -\tilde{\kappa}_{2} V_{0} - g_{2}|W_{0}|^{2}W_{0} + i\gamma_{0}W_{0},
\end{eqnarray}
where, as in the dimer case, tildes have been dropped; the coefficients of the above equations are given by:
\begin{eqnarray} \label{trimer_parameters1}
    \tilde{\kappa}_{1} \equiv \frac{\kappa}{T_{0}}\int_{0}^{T_{0}}\exp[-\Gamma(\tau)] d\tau, ~~~~~ \tilde{\kappa}_{2} \equiv \frac{\kappa}{T_{0}}
       \int_{0}^{T_{0}}\exp[\Gamma(\tau)] d\tau \nonumber \\
       \nonumber \\
    g_{1} \equiv \frac{1}{T_{0}}\int_{0}^{T_{0}}\exp[2\Gamma(\tau)] d\tau,  ~~~~~  g_{2} \equiv \frac{1}{T_{0}}
        \int_{0}^{T_{0}}\exp[-2\Gamma(\tau)] d\tau.
\end{eqnarray}
Notice that $g_{1}$ and $g_{2}$ are given by expressions identical to those
defined in the previous section. Additionally,
we will again consider the case with
$\gamma_{1}(\tau) = \gamma_{1}\cos(\tau)$ (with the constant
$\gamma_{1}$ being the modulation amplitude).

\section{Analysis of the Averaged Systems}

We will now find
solutions of the averaged effective
PT dimer and trimer models,
and investigate their stability.

\subsection{DNLS PT-symmetric dimer model}

In the averaged dimer case, we seek stationary solutions in the form:
\begin{eqnarray} \label{substitution_dimer1}
  U_{0}(T) = a\exp(-iEt), ~~~~~~~~V_{0}(t) = b\exp(-iEt)
\end{eqnarray}
where amplitudes $a, b$ are complex and frequency (or energy) $E$ is real-valued.
Moreover,
using a polar decomposition for $a$ and $b$ of the form:
\begin{eqnarray} \label{ab_dimer_form1}
 a = Ae^{i\phi_{a}}, ~~~~~~ b = Be^{i\phi_{b}},
\end{eqnarray}
we obtain the following set of real equations for $A$ and $B$:

\begin{eqnarray}
\kappa_{1} B \sin(\Delta \phi)  + \gamma_{0} A = 0, ~~~~~~ -\kappa_{2} A \sin(\Delta \phi) - \gamma_{0} B = 0 \nonumber\\
   \nonumber \\
EA = \kappa_{1} B \cos(\Delta \phi) + g_{1} A^{3}, ~~~~~~ EB = \kappa_{2} A \cos(\Delta \phi) + g_{2} B^{3},
\end{eqnarray}
where
$\Delta \phi \equiv \phi_{b} - \phi_{a}$.
The compatibility condition of the equations containing $\sin(\Delta \phi)$
yields:
\begin{eqnarray}\label{solution_dimer}
A^{2} = \frac{\kappa_{1}}{\kappa_{2}}B^{2}.
\end{eqnarray}
The above equation is then substituted into the compatibility condition of
the equations containing
$\cos(\Delta \phi)$, yielding the equation
$g_{1}\kappa_{1} = g_{2}\kappa_{2}$; the latter
is always satisfied, as seen by Eqs.~($\ref{dimer_parameters1}$). Next, we
use standard trigonometric identities to
express $A^{2}$ in terms of parameters $\kappa_{1}$,
$\kappa_{2}, \gamma_{0} ~\mbox{and}~ E$; this way, we obtain the algebraic equation
$(E - g_{1}A^{2})^{2} + {\gamma_{0}}^{2} = \kappa_{1} \kappa_{2}$, which leads to the result:
\begin{eqnarray}
      A^{2} = \frac{E \mp \sqrt{\kappa_{1} \kappa_{2} - {\gamma_{0}}^{2}}}{g_{1}}.
\end{eqnarray}

Obviously, $A^2$ is real only if
$E > \sqrt{\kappa_{1} \kappa_{2} - {\gamma_{0}}^{2}}$
or $E  >  -\sqrt{\kappa_{1} \kappa_{2} - {\gamma_{0}}^{2}}$, provided that
$\kappa_{1}\kappa_{2} - {\gamma_{0}}^{2} ~>~ 0$.
In fact, the latter inequality defines
the condition for being in the exact (and not in the broken) PT-symmetric phase.

One can also examine the stability of the stationary solutions found for the PT-symmetric effective dimer case.
Particularly, we consider the linearization ansatz on top of the stationary solutions
of Eq.~($\ref{dimer_averaged_eqns1}$)
to have the form:
\begin{eqnarray}
  U_{0}(t) = e^{-iEt}[a + pe^{\lambda t} + Pe^{\lambda^{*} t}], ~~~~~V_{0}(t) = e^{-iEt}[b + qe^{\lambda t} + Qe^{\lambda^{*} t}],
\end{eqnarray}
where the star denotes
complex conjugate. Substituting the above
ansatz into Eq.~($\ref{dimer_averaged_eqns1}$) and linearizing
in $p, P, q ~\mbox{and}~ Q$,
we obtain the eigenvalue problem:
    \begin{eqnarray}
       {\bf A}{\bf X} = i\lambda {\bf X},
       \label{ep}
    \end{eqnarray}
    where ${\bf X} = (p, ~-P^{*}, ~q, ~-Q^{*})^{T}$ and the $4\times4$ matrix ${\bf A}$ has elements $a_{ij}$ given by:
\begin{eqnarray}
    a_{11} = -E + 2g_{1}|a|^{2} + i\gamma_{0} ~~~a_{12} = g_{1}a^{2}, ~~~a_{13} = \kappa_{1}, ~~~a_{14} = 0
           \nonumber\\
             \nonumber \\
    a_{21} = -g_{1}(a^{*})^{2}, ~~~a_{22} = E - 2g_{1}|a|^{2} + i\gamma_{0}, ~~~a_{23} = 0,  ~~~a_{24} = -\kappa_{1}
           \nonumber\\
             \nonumber \\
     a_{31} = \kappa_{2}, ~~~a_{32} = 0, ~~~a_{33} = -E + 2g_{2}|b|^{2} - i\gamma_{0}, ~~~a_{34} = g_{2}b^{2}
           \nonumber\\
             \nonumber \\
     a_{41} = 0, ~~~a_{42} = -\kappa_{2}, ~~~a_{43} = -g_{2}(b^{*})^{2}, ~~~a_{44} = E - 2g_{2}|b|^{2} - i\gamma_{0}.
\end{eqnarray}
Upon substituting
the parameters characterizing the solutions of the
PT-symmetric dimer model into $a_{ij}$, and solving the eigenvalue problem (\ref{ep}), one can then
find the eigenvalues $\lambda$, which
determine the spectral stability of the corresponding nonlinear solutions:
the existence of eigenvalues with positive real part,
$\lambda_{r} > 0$, amounts to a dynamical instability of the relevant solution, while in the case
where all the eigenvalues have $\lambda_{r} \le 0$,
the solution is linearly stable. We will offer more details on the
specifics of the linearization analysis in the numerical section, for
the particular choice of cosinusoidal dependence of $\gamma$ on time considered herein.

\subsection{DNLS PT-symmetric trimer model}

First, we rewrite Eqs.~(\ref{trimer_averaged_eqns1}) in the following form:

\begin{eqnarray} \label{trimer_averaged_eqns2}
  i\frac{\partial U_{0}}{\partial t} = k_{1} V_{0} + g_{1}|U_{0}|^{2}U_{0} + i\gamma_{0}U_{0},
      \nonumber \\
   i\frac{\partial V_{0}}{\partial t} = k_{2} U_{0} + k_{1}W_{0}  + |V_{0}|^{2}V_{0},
      \nonumber \\
   i\frac{\partial W_{0}}{\partial t} = k_{2}V_{0} + g_{2}|W_{0}|^{2}W_{0} - i\gamma_{0}W_{0}.
\end{eqnarray}
where the averaged coefficients $k_{1} \equiv -\tilde{\kappa}_{1}$, $k_{2} \equiv - \tilde{\kappa}_{2}$,
$g_{1} ~\mbox{and}~ g_{2}$ are given in Eq.~($\ref{trimer_parameters1}$).
We again seek stationary solutions of the form:
\begin{eqnarray} \label{substitution_trimer1}
  U_{0}(t) = a\exp(-iEt), ~~~~~~V_{0}(t) = b\exp(-iEt), ~~~~~~W_{0}(t) = c\exp(-iEt)
\end{eqnarray}
where $E$ is real-valued and the complex amplitudes $a, b$ and $c$ are decomposed as:
\begin{eqnarray} \label{polar_trimer1}
  a = Ae^{i\phi_{a}}, ~~~~b = Be^{i\phi_{b}}, ~~~~c = Ce^{i\phi_{c}}.
\end{eqnarray}

Substituting the above expressions into Eqs.~($\ref{trimer_averaged_eqns2}$) we obtain
the following system for $A$, $B$ and $C$:
\begin{eqnarray} \label{ABC_real_eqns1}
  k_{1}B \sin(\Delta \phi_{1}) + \gamma_{0}A = 0, ~~~~~ k_{2}B \sin(\Delta \phi_{2}) - \gamma_{0}C = 0, \hspace*{1.0in} \nonumber \\
     \nonumber \\
 EA = k_{1}B \cos(\Delta \phi_{1}) + g_{1}A^{3}, ~~~~~ EC = k_{2}B \cos(\Delta \phi_{2}) + g_{2}C^{3}, \hspace*{0.8in} \nonumber \\
      \nonumber \\
 EB = k_{2}A \cos(\Delta \phi_{1}) + k_{1}C \cos(\Delta \phi_{2}) + B^{3}, ~~~~~ -k_{2}A \sin(\Delta \phi_{1}) - k_{1}C \sin(\Delta \phi_{2}) = 0,
\end{eqnarray}
where $\Delta \phi_{1} \equiv \phi_{b} - \phi_{a}$ and $\Delta \phi_{2} \equiv \phi_{b} - \phi_{c}$.
We determine nontrivial solutions for
$A, B$ and $C$ by solving the first four equations in Eq.~($\ref{ABC_real_eqns1}$) for $\sin(\Delta \phi_{1})$,
$\sin(\Delta \phi_{2})$, $\cos(\Delta \phi_{1})$ and
$\cos(\Delta \phi_{2})$, and then plugging these results into the last two equations in Eq.~($\ref{ABC_real_eqns1}$).
This way, we derive the following two consistency conditions:
\begin{eqnarray} \label{ABC_algebraic2}
  k_{1}k_{2}B^{4} - k_{1}k_{2}EB^{2} + E(k_{2}^{2}A^{2} + k_{1}^{2}C^{2}) - (k_{2}^{2}g_{1}A^{4}
  + k_{1}^{2}g_{2}C^{4}) = 0,
  ~~~~~ \gamma_{0}(k_{1}^{2}C^{2} - k_{2}^{2}A^{2}) = 0.
\end{eqnarray}
The second equation in ($\ref{ABC_algebraic2}$) leads to a relation connecting
$C^{2}$ and $A^{2}$, namely $C^{2}=(k_2/k_1)^2 A^2$, which
must be imposed to satisfy the first of Eqs.~($\ref{ABC_algebraic2}$). Using this
relation, and the first two sets of equations in ($\ref{ABC_real_eqns1}$), we find:
\begin{eqnarray}
  \cos(\Delta \phi_{1}) = \frac{A(E - g_{1}A^{2})}{k_{1}B}, ~~~~~~
    \cos(\Delta \phi_{2}) = \pm \frac{A\left(E - g_{2}(\frac{k_{2}}{k_{1}})^{2}A^{2}\right)}{k_{1}B}, 
        \nonumber \\
   \sin(\Delta \phi_{1}) = -\frac{\gamma_{0}A}{k_{1}B}, ~~~~~~ \sin(\Delta \phi_{2}) = \pm \frac{\gamma_{0}A}{k_{1}B}. \hspace*{0.75in}
\end{eqnarray}
To this end, we use
trigonometric identities
to finally connect $A$ and $B$ through the algebraic conditions:
\begin{eqnarray}
  {g_{1}}^{2}A^{6} - 2Eg_{1}A^{4} + (E^{2} + {\gamma_{0}}^{2})A^{2} - {k_{1}}^{2}B^{2} = 0, \hspace*{0.5in} 
      \nonumber \\
     {g_{2}}^{2}\left(\frac{k_{2}}{k_{1}} \right)^{4}A^{6} - 2Eg_{2}\left(\frac{k_{2}}{k_{1}}\right)^{2}A^{4} + (E^{2} + {\gamma_{0}}^{2})A^{2}
      - {k_{1}}^{2}B^{2} = 0.
\end{eqnarray}
These equations are consistent (i.e., reduce to a single equation) if one requires $g_{1} = g_{2}(k_{2}/k_{1})^{2}$.
Using this requirement,
along with $C^{2} = (k_{2}/k_{1})^{2}A^{2}$ and Eq.~($\ref{ABC_algebraic2}$), we derive
two equations that can be used to determine $A$ and $B$ explicitly:
\begin{eqnarray} \label{AB_explicit_eqns}
  k_{1}k_{2}B^{4} - k_{1}k_{2}EB^{2} + 2E{k_{2}}^{2}A^{2} - 2g_{1}{k_{2}}^{2}A^{4} = 0, ~~~~~~
    {g_{1}}^{2}A^{6} - 2Eg_{1}A^{4} + (E^{2} + {\gamma_{0}}^{2})A^{2} - {k_{1}}^{2}B^{2} = 0,
\end{eqnarray}
where $C^{2} = (k_{2}/k_{1})^{2}A^{2}$ and $g_{1} = g_{2}(k_{2}/k_{1})^{2}$. One can then  solve Eqs.~($\ref{AB_explicit_eqns}$) for $A$ and $B$
in terms of parameters $g_{1}, k_{1}, k_{2}, E, \gamma_{0} ~\mbox{and}~ g_{2}$.

In a similar manner to the dimer case, once the relevant stationary
states are obtained, one can examine the stability of the stationary solutions found for the PT-symmetric effective trimer  case. We consider solutions
of Eqs.~($\ref{trimer_averaged_eqns2}$) of the form:
\begin{eqnarray}
  U_{0}(t) = e^{-iEt}[a + pe^{\lambda t} + Pe^{\lambda^{*} t}], ~~~~~~V_{0}(t) = e^{-iEt}[b + qe^{\lambda t} + Qe^{\lambda^{*} t}], \nonumber \\
         \nonumber \\
  W_{0}(t) = e^{-iEt}[c + re^{\lambda t} + Re^{\lambda^{*} t}]. \hspace*{1.0in}
\end{eqnarray}
Substituting this ansatz into Eq.~($\ref{trimer_averaged_eqns2}$) and
linearizing in $p, P, q, Q, r ~\mbox{and}~ R$, we end up with
the eigenvalue problem:
    \begin{eqnarray}
       {\bf A}{\bf Y} = i\lambda {\bf Y}
       \label{ep2},
    \end{eqnarray}
where ${\bf Y} = (p, ~-P^{*}, ~q, ~-Q^{*}, ~r, ~-R^{*})^{T}$ and the $6\times6$ stability matrix ${\bf A}$
has elements $a_{ij}$ which are now given by:

 \begin{eqnarray}
   a_{11} = -E + 2g_{1}|a|^{2} + i\gamma_{0}, ~~~a_{12} = g_{1}a^{2}, ~~~a_{13} = k_{1}, ~~~a_{14} = 0, ~~~a_{15} = 0, ~~~a_{16} = 0,
     \nonumber \\
      \nonumber \\
   a_{21} = -g_{1}(a^{*})^{2}, ~~~a_{22} = E - 2g_{1}|a|^{2} + i\gamma_{0}, ~~~a_{23} = 0, ~~~a_{24} = -k_{1}, ~~~a_{25} = 0, ~~~a_{26} = 0,
      \nonumber \\
      \nonumber \\
   a_{31} = k_{2}, ~~~a_{32} = 0, ~~~a_{33} = -E + 2|b|^{2}, ~~~a_{34} = b^{2}, ~~~a_{35} = k_{1}, ~~~a_{36} = 0,
      \nonumber \\
      \nonumber \\
   a_{41} = 0, ~~~a_{42} = -k_{2} , ~~~a_{43} = -(b^{*})^{2}, ~~~a_{44} = E - 2|b|^{2}, ~~~a_{45} = 0, ~~~a_{46} = -k_{1},
      \nonumber \\
      \nonumber \\
  a_{51} = 0, ~~~a_{52} = 0, ~~~a_{53} = k_{2}, ~~~a_{54} = 0, ~~~a_{55} = -E + 2g_{2}|c|^{2} - i\gamma_{0}, ~~~a_{56} = g_{2}c^{2},
      \nonumber \\
      \nonumber \\
  a_{61} = 0, ~~~a_{62} = 0, ~~~a_{63} = 0, ~~~a_{64} = -k_{2}, ~~~a_{65} = -g_{2}(c^{*})^{2}, ~~~a_{66} = E - 2g_{2}|c|^{2} - i\gamma_{0}.
 \end{eqnarray}
As before, substitution of the parameters of the solutions in $a_{ij}$, and solution of (\ref{ep2}) will lead to the
eigenvalues that
determine the spectral stability of the corresponding nonlinear solutions.
Once again, the details of the linear stability properties
will be explored in the upcoming numerical section.

\section{Numerical results for the modulated system and comparison to the averaged models}

We show below the results of the numerical analysis of the full non-autonomous
system of Eqs.~(\ref{dnls1}) and (\ref{dnls_trimer1}) and compare them
to those of the averaged equations derived above. In order to simplify the notation, we denote $y\equiv\{u,v\}$ for the dimer and $y\equiv\{u,v,w\}$ for the trimer. In order to get exact periodic orbits of period $T_b=2\pi/\omega$, we use
\begin{equation}
    y(t)=\exp(-iEt)x(t),
\end{equation}
and $x(t)$ is found by means of a shooting method,
which is based on finding
fixed points of the map $x(0)\rightarrow x(T_b)$. The stability of the periodic orbit is obtained by means of a Floquet  method, which identifies the relevant Floquet multipliers; see e.g.~\cite{shooting} for a relevant discussion and several application examples. In order to apply Floquet method, a small perturbation $\xi(t)$ is added to a given solution $x(t)$, and the stability properties of the solutions are given by the spectrum of the Floquet operator whose matrix representation is the {\em monodromy} matrix $\mathcal{M}$.
The monodromy matrix eigenvalues $\Lambda$ are dubbed as Floquet multipliers. This operator is real, which implies that there is  always a pair of multipliers at $1$ (corresponding to the so-called phase and growth modes) and that the eigenvalues come in pairs $\{\Lambda,\Lambda^*\}$.

In order to preserve the PT symmetry of the system, $\gamma_1(t)$ must be even in time. In that light, as indicated above, we have chosen
\begin{equation}
    \gamma_1(t)=\gamma_1\cos(\omega t),
\end{equation}
i.e., $\epsilon\equiv1/\omega$ and $\tau=t/\epsilon=\omega t$ in Eqs.~(\ref{dnls1}) and (\ref{dnls_trimer1});
consequently, if the period of $\gamma_1$ is
$T_0=2\pi$, this implies that
\begin{equation}
   \frac{1}{T_0}\int_{0}^{T_{0}}\exp[\pm\beta\Gamma(\tau)] d\tau=I_0(\beta\gamma_1),
\end{equation}
with $\beta$ being an arbitrary real number, and $I_0$ being the
zeroth-order modified Bessel function of the first kind. Given that $I_0(x)=I_0(-x)$, we can write $\kappa_1=\kappa_2=\kappa'\equiv\kappa I_0(\beta_1\gamma_1)$, $g_1=g_2=g'\equiv I_0(\beta_2\gamma_1)$,
with the values of $\beta_1$ and $\beta_2$ depending on the particular oligomer we are dealing with.

We have compared below the numerical results of the full non-autonomous
problem with the predictions from the averaged system when the modulation amplitude is $\gamma_1=1$
(results for other values of $\gamma_1$ were also considered, and qualitatively similar results were obtained).
We have analyzed a fast modulation $\omega=1000$ and a considerably
slower one of $\omega=20$. In the former case,
the agreement is excellent, i.e., the curves from the
non-autonomous and the averaged problem cannot be distinguished; for this reason,
the results for that case will not be shown.
Thus, below, we will restrict our numerical presentation
to the slower modulated case of $\omega=20$.

The quantities compared between the effective averaged solution
properties and those of the non-autonomous system
are the Floquet multipliers,
which are related to the stability eigenvalues $\lambda$ of the averaged
system as:
\begin{equation}
    \Lambda=\exp(2\pi\lambda/\omega),
\end{equation}
and the averaged $\ell^2$ norm (or average power)
of the corresponding vector $y$,
defined as
\begin{equation}
    <N>=\frac{1}{T_{b}}\int_{0}^{T_{b}} |y(t)|^2 dt.
\end{equation}
Note that the predictions made for the averaged system are able only to
obtain $Y\equiv\{U,V\}$ or $Y\equiv\{U,V,W\}$, so the change of variables in (\ref{dnls_var1}) or (\ref{dnls_trimer_var1}) must be taken into account in order to
find $y(t)$.

\subsection{DNLS PT-symmetric dimer model}

In this case, $\kappa'=\kappa q$, and $g'=q$ with $q=I_0(2\gamma_1)$, and, consequently, (\ref{solution_dimer}) yields for the averaged system:
\begin{equation}
    A^2=B^2=\frac{E\mp\sqrt{\kappa^2 q^2-\gamma_0^2}}{q},
\end{equation}
with the $\mp$ sign
corresponding, respectively, to the symmetric (S) and anti-symmetric (A) solution (of the Hamiltonian limit). Consequently, there
is a saddle-center bifurcation at $\gamma_0=\kappa q$. Above this value,
the amplitudes become imaginary and the relevant analytical
solutions of the effective system do not exist. This is precisely,
the nonlinear analogue of the PT phase transition; note that the latter, especially
in the case of the dimer,
coincides with the linear PT phase transition.
A remarkable feature that we observe in this context
is that, since $q > 1$, the critical point for both the linear
and the nonlinear PT symmetry-breaking transition will be increased,
hence the region of gain/loss parameters $\gamma_0$
corresponding to an exact PT-symmetric
phase will be expanded (possibly quite considerably and, in any case,
controllably so) with respect to the unmodulated case.

Interestingly, in the case of the dimer, following the analysis of Ref.~\cite{pgk},
the linear stability eigenvalues can be analytically found for the effective dimer as:
\begin{equation}
    \lambda=\pm 2i\sqrt{2(\kappa^2 q^2-\gamma_0^2)\mp E\sqrt{\kappa^2 q^2-\gamma_0^2}}.
\end{equation}
The numerical analysis of the modulated system is done by choosing $\kappa=1$ and $E=3$. As explained above, $\gamma_1=1$, with $\omega=20$, yielding $q=I_0(2)\approx2.2796$. If $E>q$, the S and A solutions exist at $\gamma_0=0$.

\begin{figure}
\begin{center}
\begin{tabular}{cc}
    \includegraphics[scale=0.4]{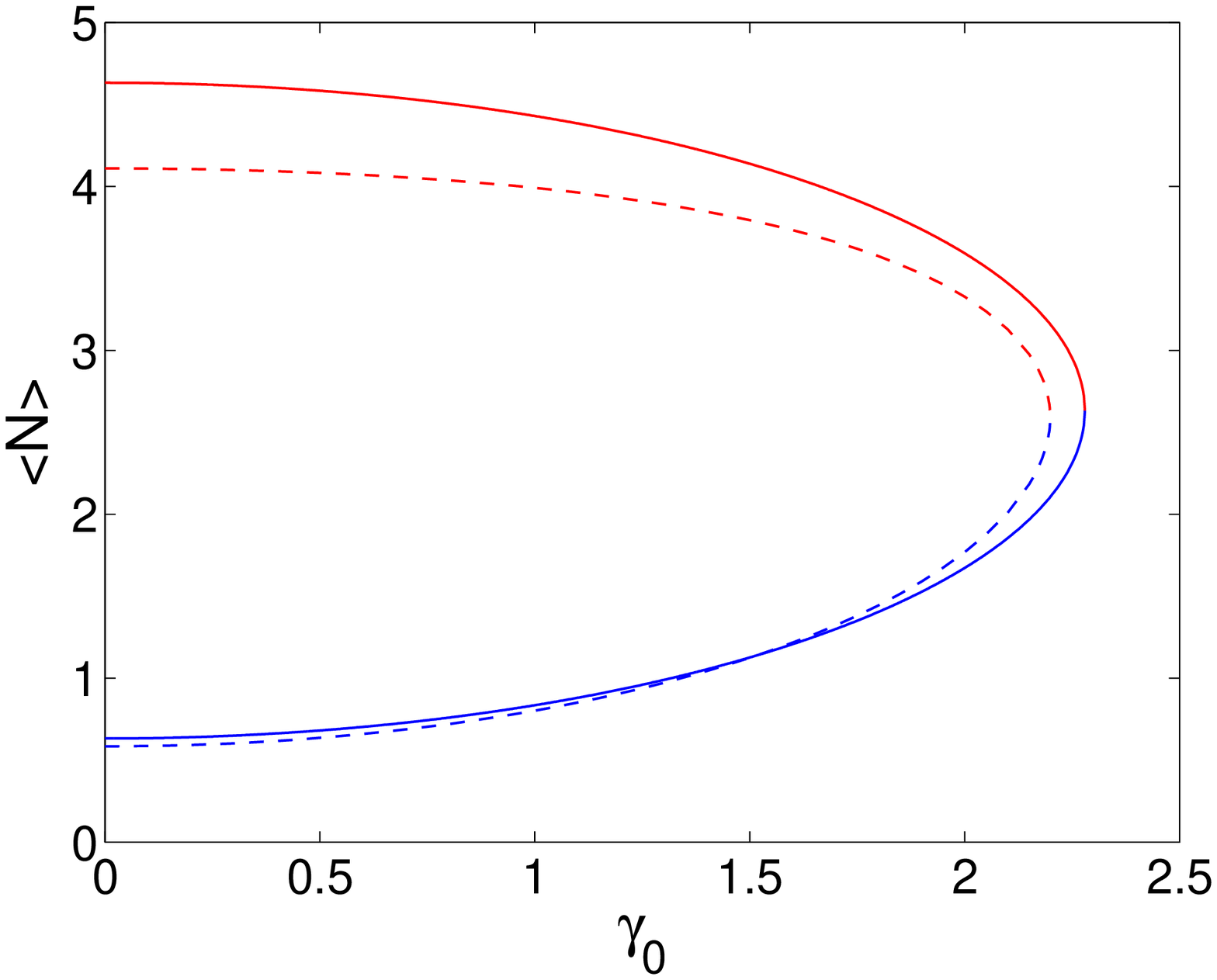} &
    \includegraphics[scale=0.4]{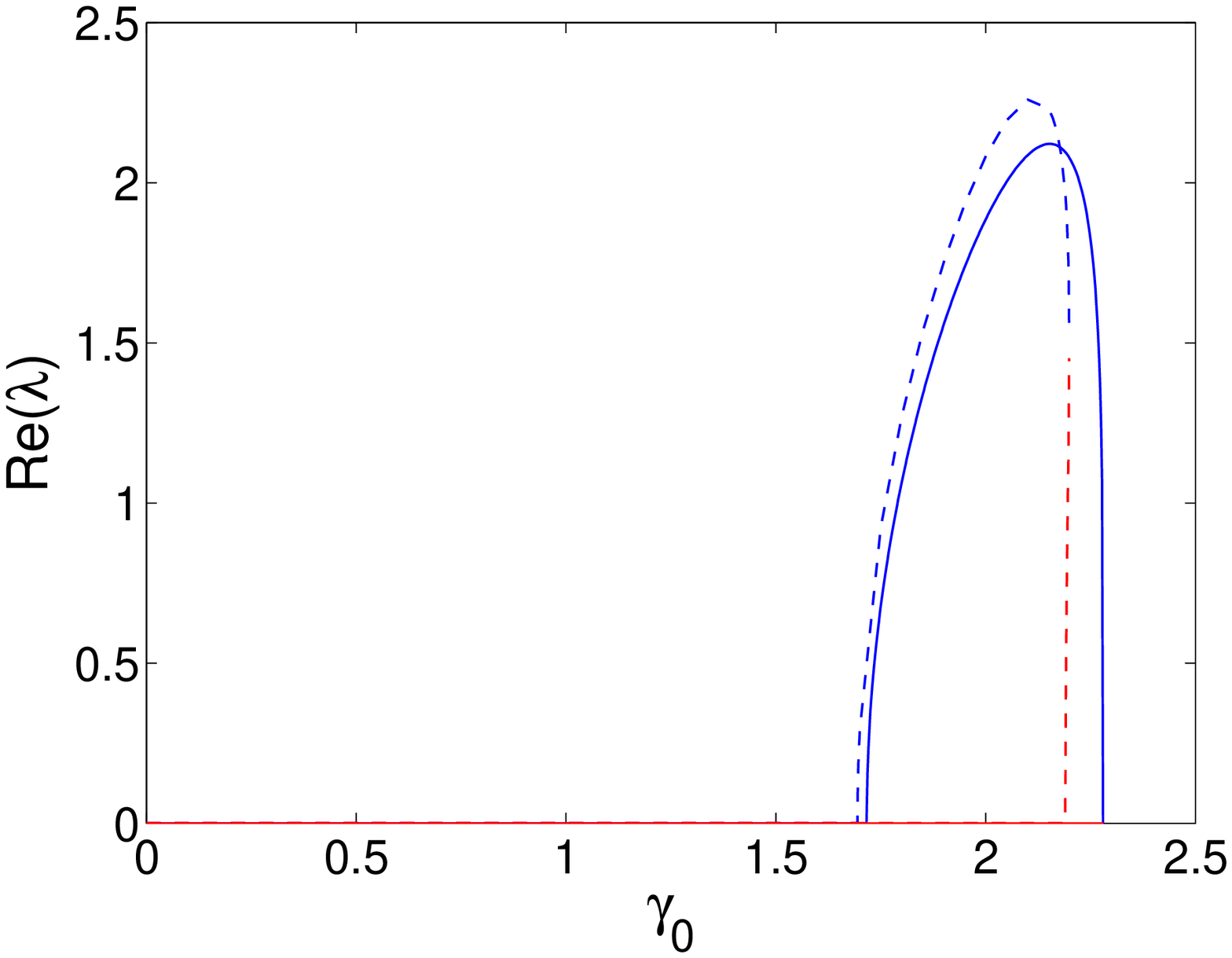} \\
\end{tabular}
\caption{(Color online) Averaged norm (left panel) and real part of the stability eigenvalues (right panel)
for a dimer with $\kappa=1$, $E=3$ and $\omega=\gamma_1=20$.
Solid (dashed) lines represent the values for the averaged (modulated) system, whereas blue (red) corresponds
to the S (A) solution.
The designation of S (symmetric) or A (anti-symmetric) corresponds
to the $\gamma_0=0$ Hamiltonian limit of the problem where $\Delta \phi=0$
or $\pi$, respectively.} \label{fig:dimer}
\end{center}
\end{figure}

Figure~\ref{fig:dimer} shows both the averaged
norm and the real part of the stability eigenvalues, together with the
predicted values by the averaged equations. We can observe that the
bifurcation designated as the nonlinear
analogue of the PT-phase transition (leading to the collision
and disappearance of the --former-- symmetric and anti-symmetric
solutions of the $\gamma_0=0$ limit) takes place only slightly
earlier in the
non-autonomous system (at $\gamma_0=2.1989$). Regarding the stability,
it is predicted in the averaged system that the A solution becomes unstable
for $\gamma_0 = \sqrt{\kappa^2 q^2-E^2/4}\approx1.7165$;
in the modulated system,
this bifurcation takes place around $\gamma_0=1.6945$ (i.e., again
at a very proximal value). Additionally, and quite interestingly,
even the S solution may become unstable close to the transition
point i.e, for $\gamma_0 > 2.1899$. This feature is not captured
by the averaged equations but also only appears to be a very weak
and hence not particularly physically significant effect.
Note that the fact that both pairs of multipliers for the A and S solutions come
in towards the bifurcation point from the unstable side has been observed recently
in a PT-symmetric Klein--Gordon dimer \cite{ptkg}.

\begin{figure}
\begin{center}
\begin{tabular}{cc}
    \includegraphics[scale=0.4]{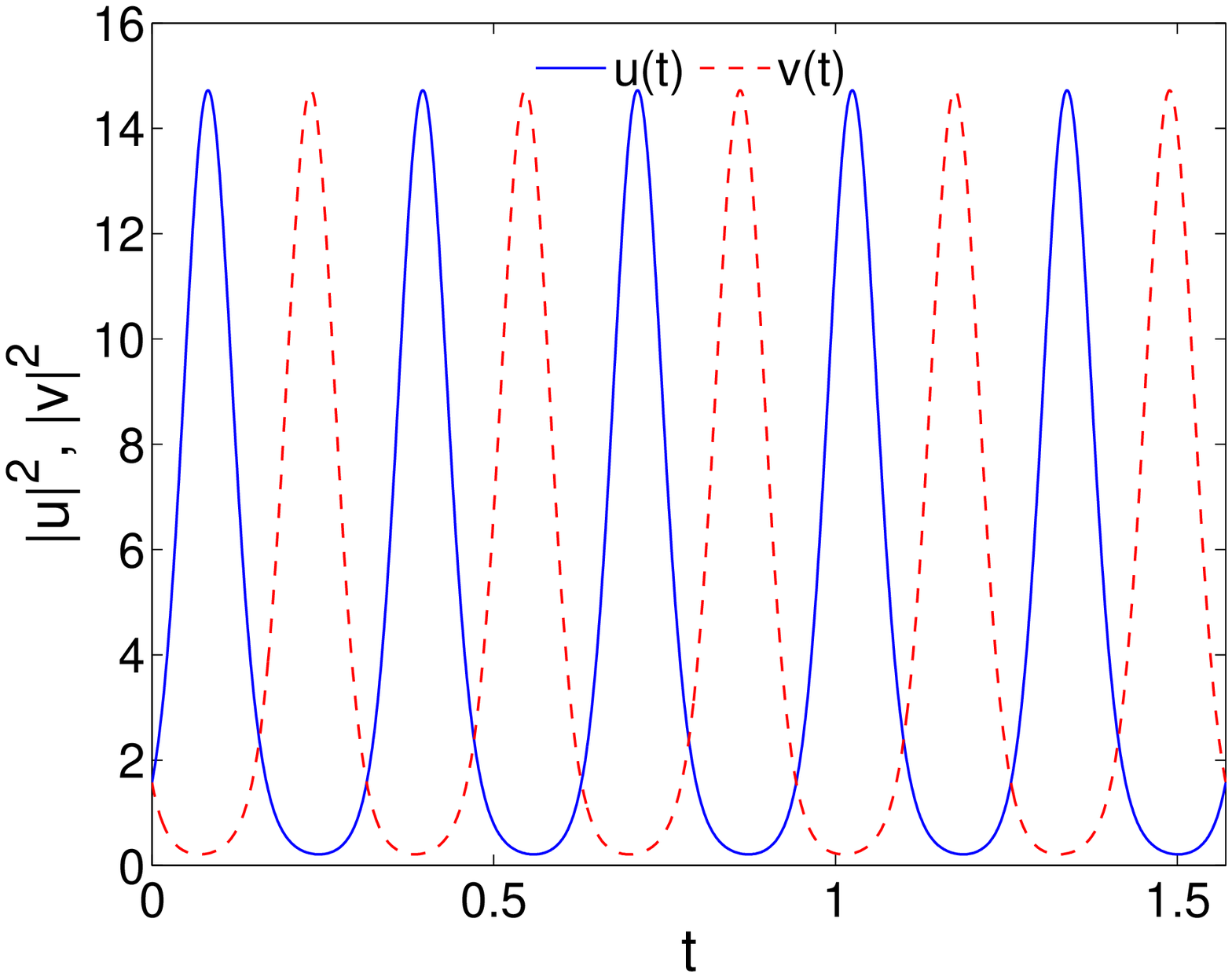} &
    \includegraphics[scale=0.4]{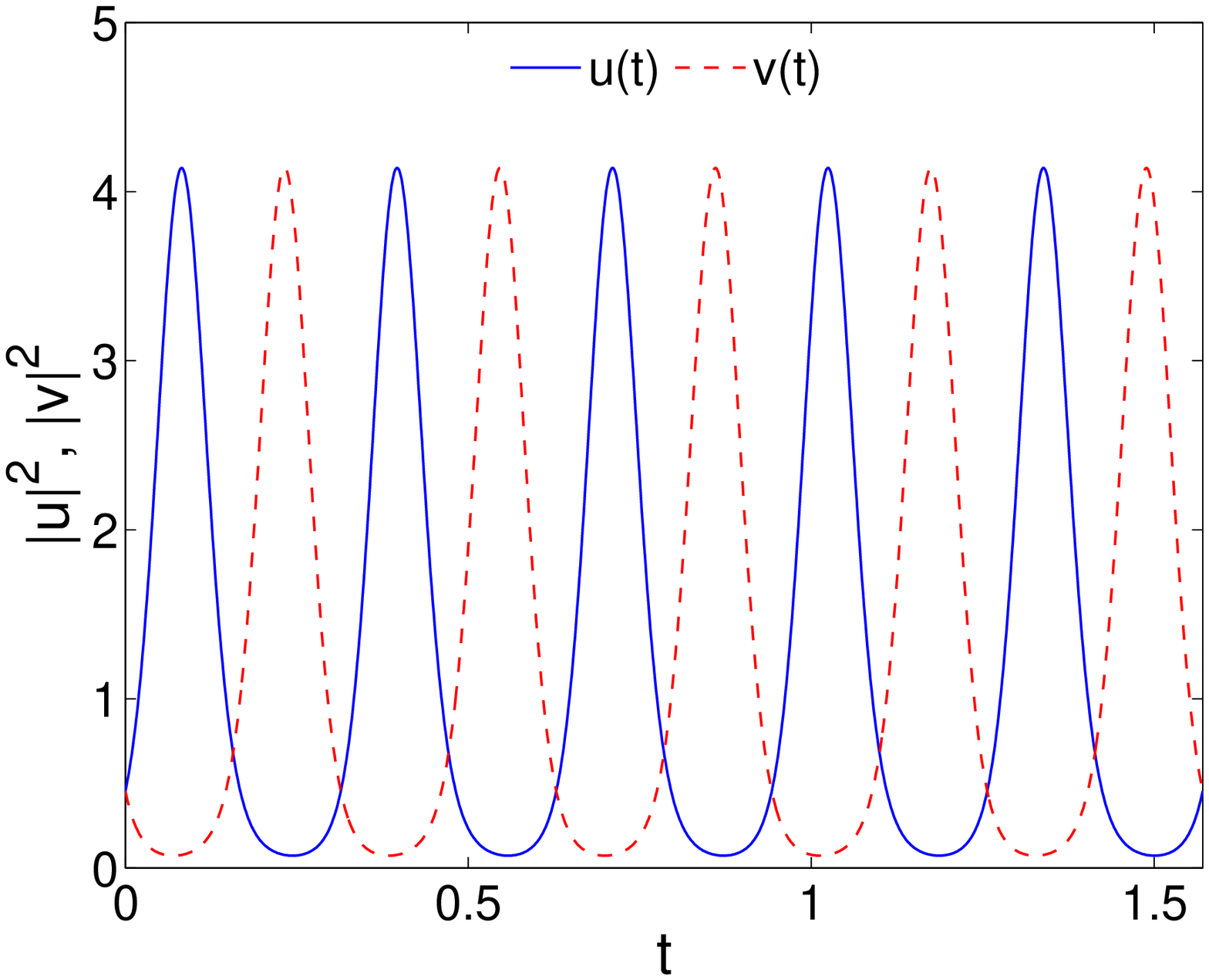} \\
    \includegraphics[scale=0.4]{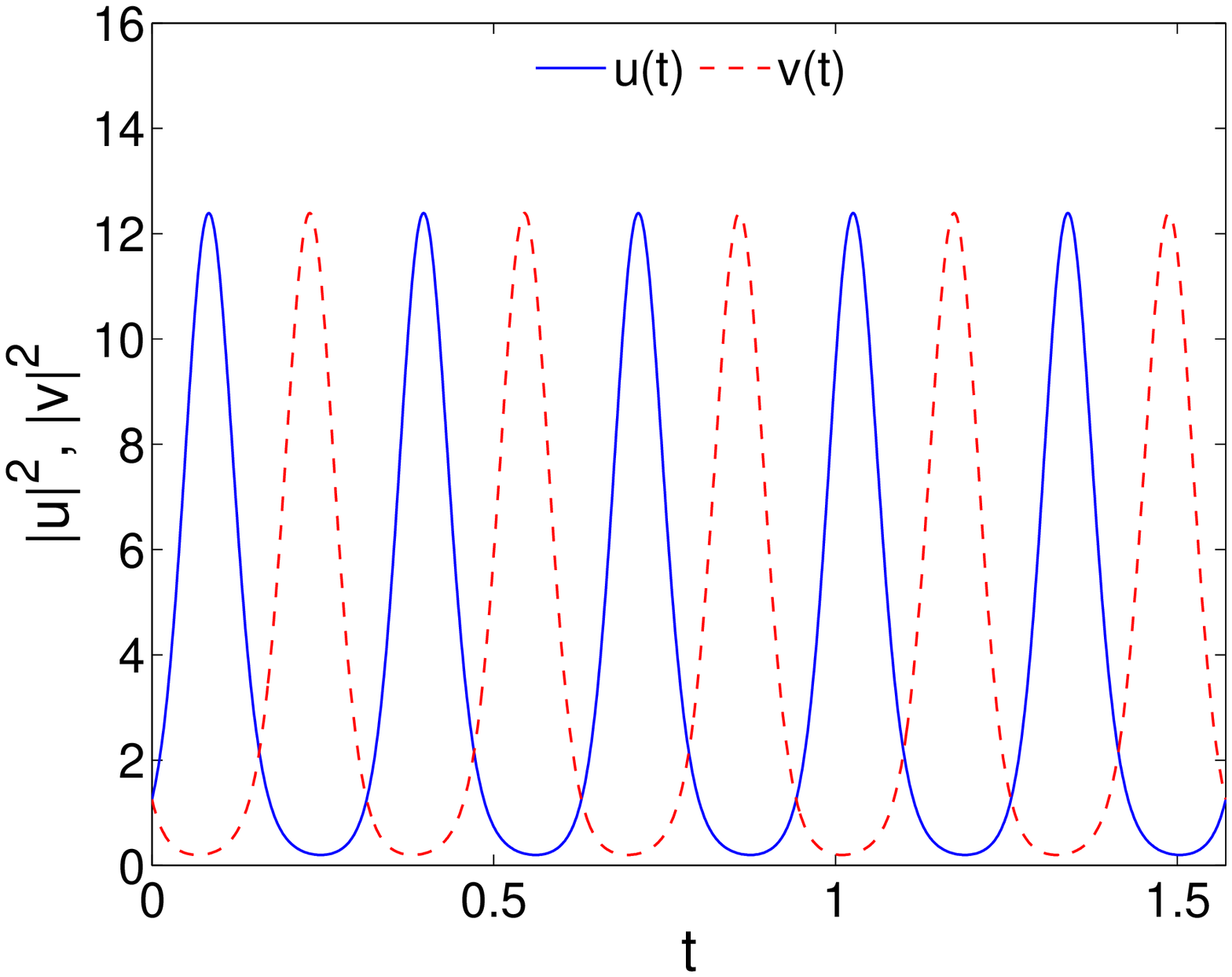} &
    \includegraphics[scale=0.4]{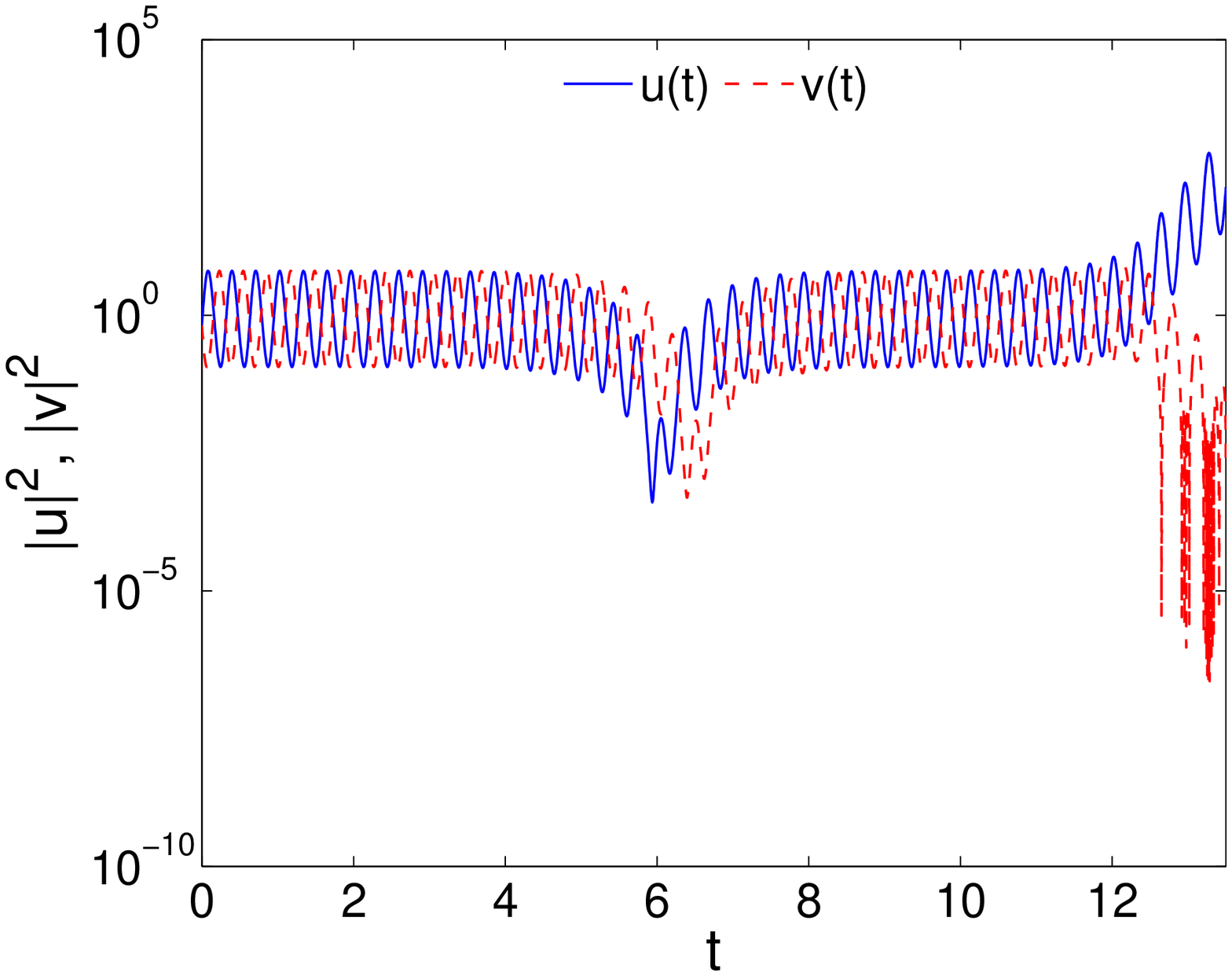} \\
    \includegraphics[scale=0.4]{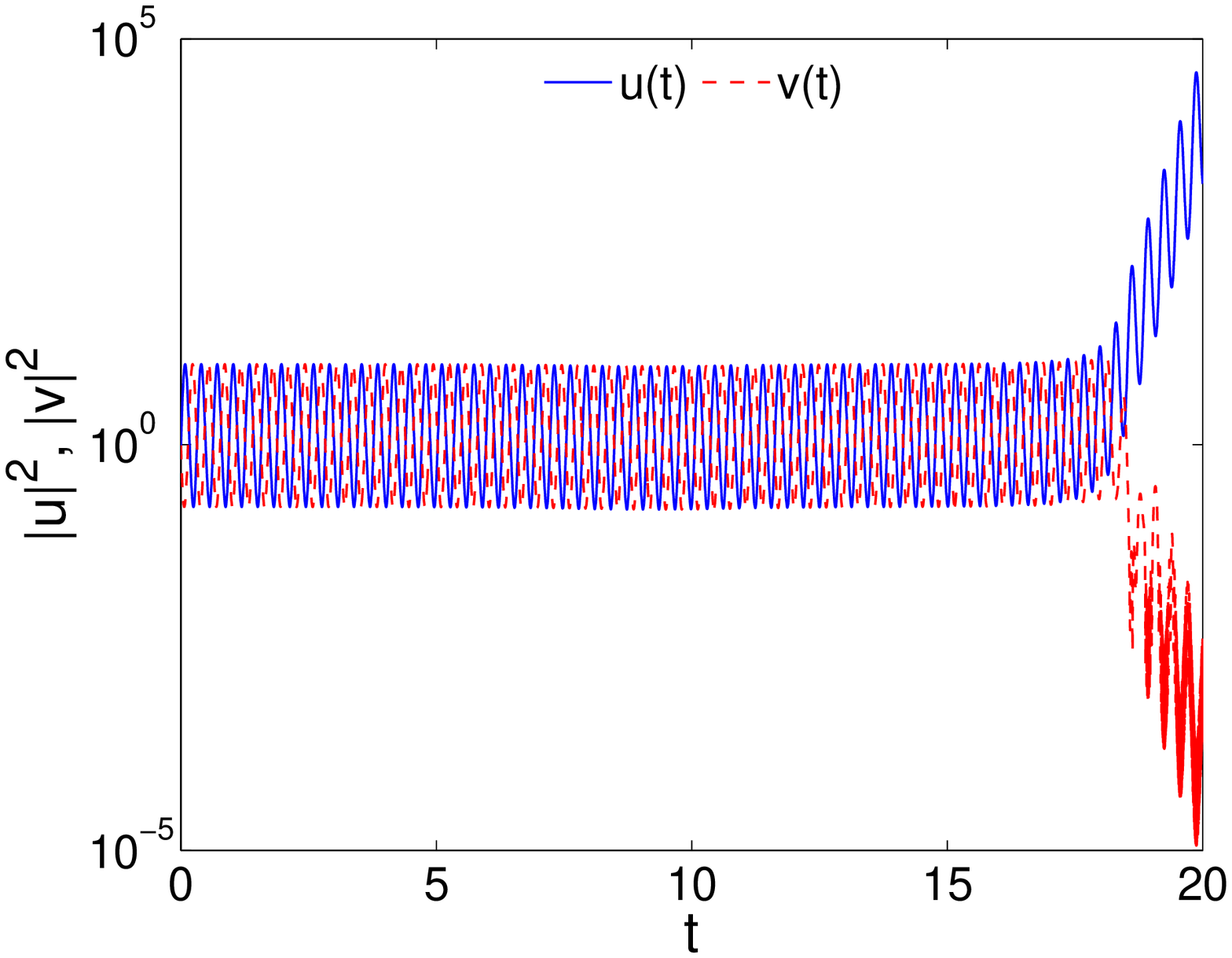} &
    \includegraphics[scale=0.4]{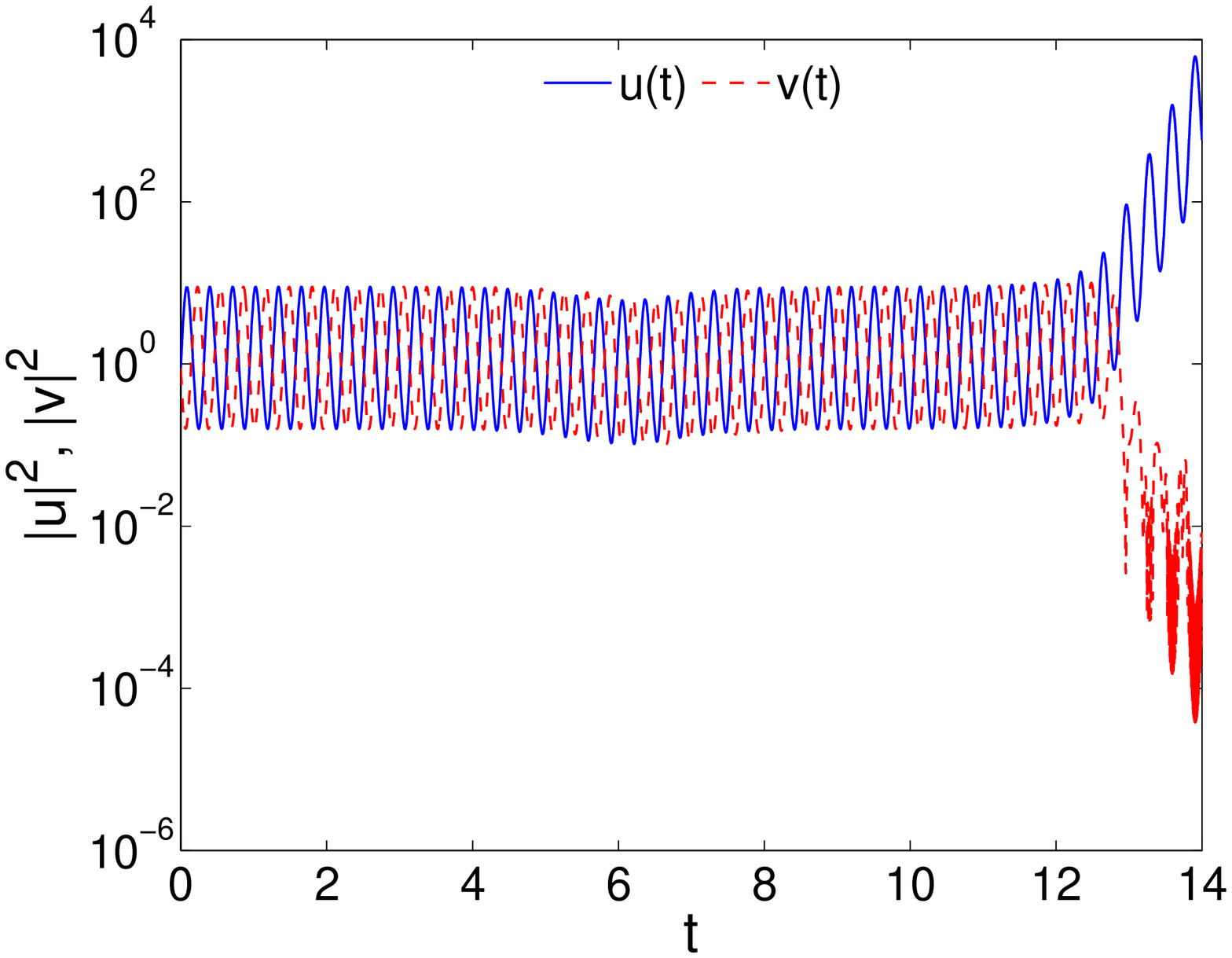} \\
    \includegraphics[scale=0.4]{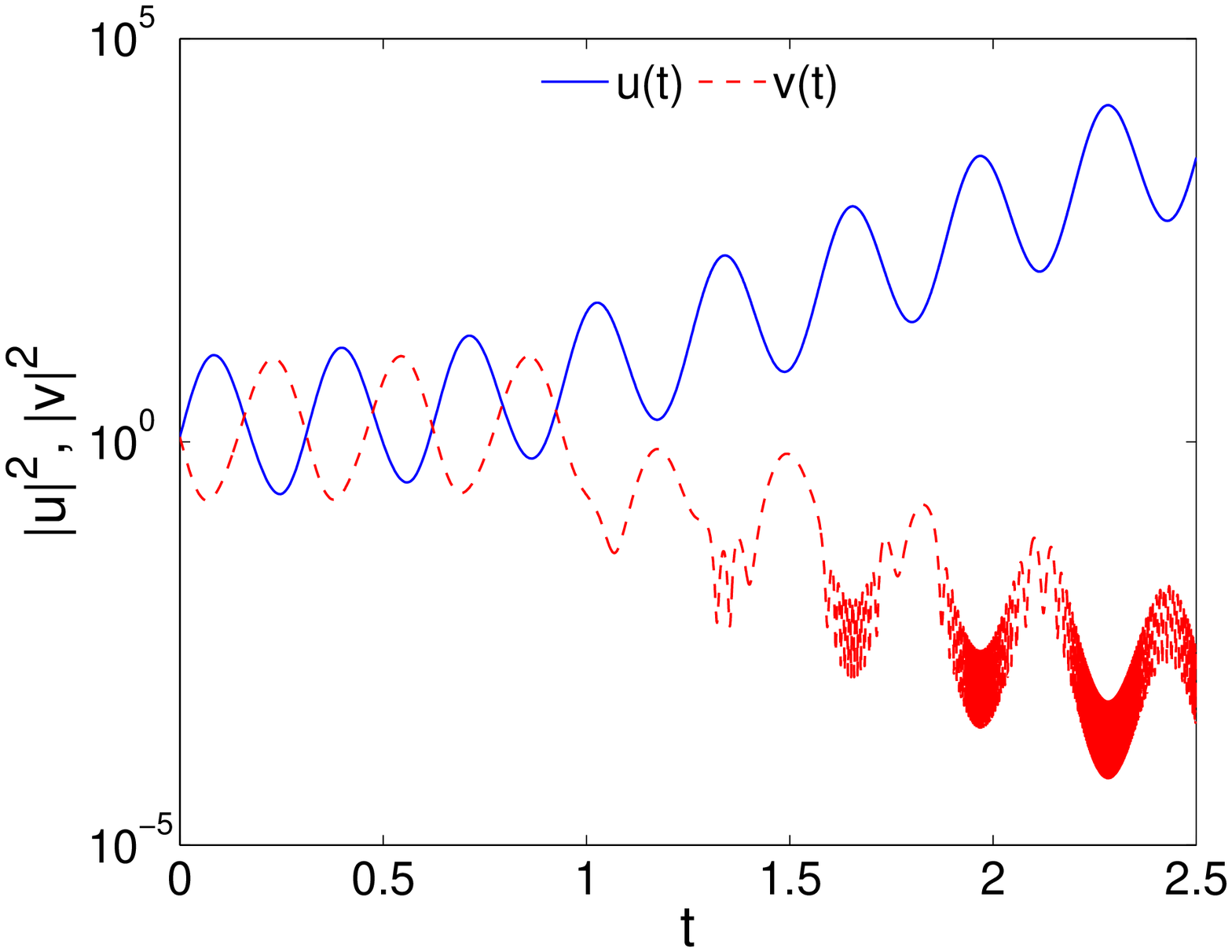} &
    \includegraphics[scale=0.4]{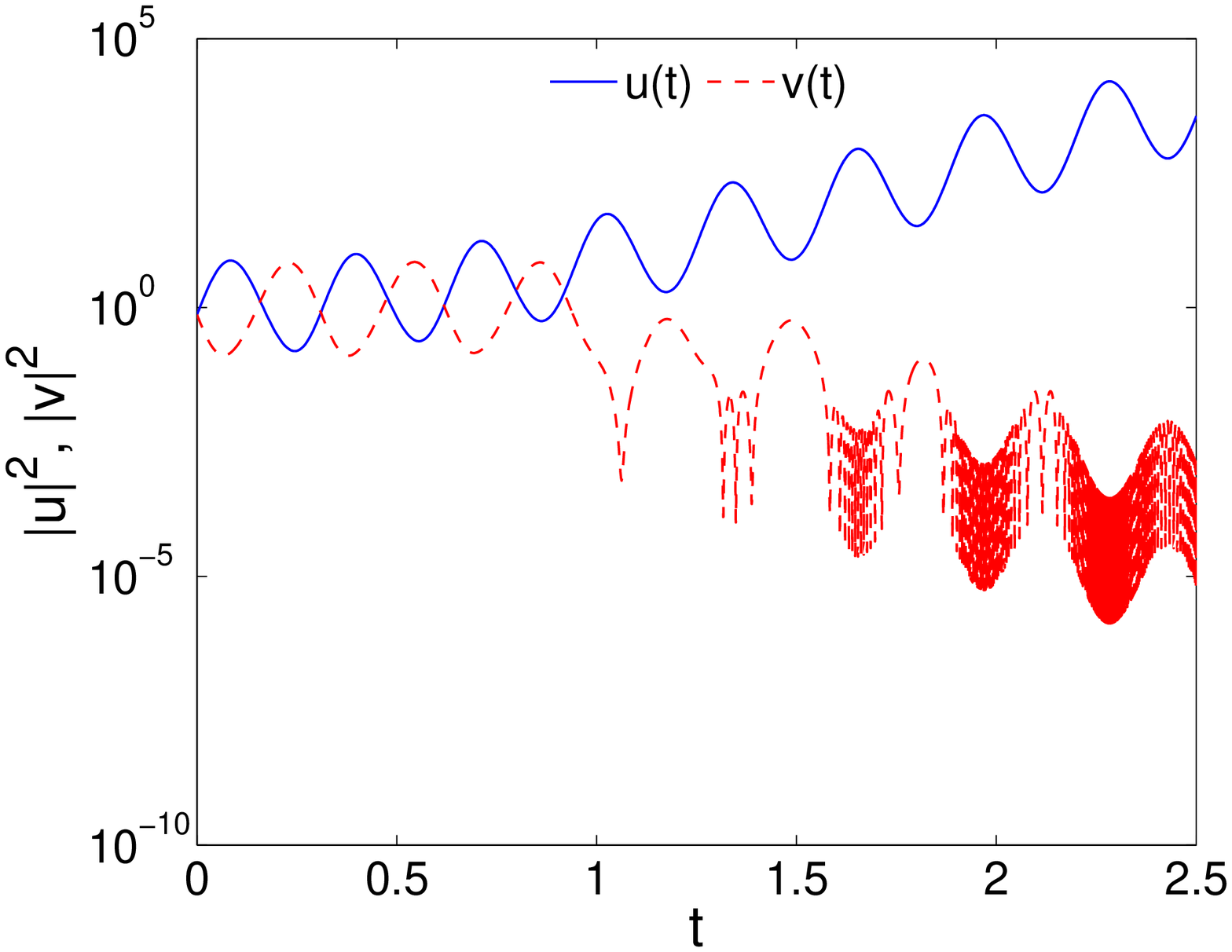} \\
\end{tabular}
\caption{(Color online) Dynamical evolution of former (at $\gamma_0=0$) A solutions (left panels) and S solutions
(right panels) for the modulated dimer in the case $\kappa=1$, $E=3$, $\gamma_1=1$ and $\omega=20$. Top panels
correspond to a stable evolution at $\gamma_0=1.5$ (which is already beyond the PT-phase transition critical point for $\gamma_0$ in the case of $\gamma_1=0$); panels of the second row show the stable (unstable) evolution
for the A (S) solution at $\gamma_0=2$, whereas the third row displays the unstable evolution of solutions at
$\gamma_0=2.195$. Finally, bottom panels correspond to the evolution at $\gamma_0=2.3$ (i.e. past the PT symmetry breaking
bifurcation) taking as initial condition the solution at $\gamma_0=2.1$.}
\label{fig:dimer_dynamics}
\end{center}
\end{figure}

Figure~\ref{fig:dimer_dynamics} shows the dynamical evolution of stable and unstable (A and S) solutions.
The top panels illustrate a case example of stable oscillations for
$\gamma_0=1.5$. Notice, however, that this value of the gain/loss
parameter is already above the critical one in the absence of
modulation, clearly showcasing the extension of the PT-symmetric
regime due to the presence of the modulation. Here, the elements of both
branches execute stable periodic motion. In the second row, for
$\gamma_0=2$, the former
anti-symmetric
oscillation remains stable, but the former
symmetric
one is in its regime of instability, thus giving rise to a modulated form
of growth, whereby the gain site grows indefinitely while the lossy
site ultimately approaches a vanishing amplitude.
The third row shows the evolution for $\gamma_0=2.195$, a value for which both A and S solutions are unstable.
At the {\it modified} threshold (fourth row) of the PT-symmetry breaking, both
branches are sensitive to perturbations and can give rise to growth of
one node and decay of the other. This type of behaviour is also
generically observed to be relevant for initial data beyond the PT-phase transition
threshold, as is shown in the bottom row.

\subsection{DNLS PT-symmetric trimer model}

We now turn to the analysis of the trimer case, where
$\kappa'=\kappa I_0(\gamma_1)$ and $g'=I_0(2\gamma_1)$. In this case,
the linear stability eigenvalues of the averaged system are unfortunately
not available analytically and are, instead, found by numerical
diagonalization. In the numerics, we have chosen the same parameters as in the
dimer case except for $E=1$.

In agreement with what has been reported earlier for
Sch{\"o}dinger trimers without time-modulation~\cite{pgk,kipnew},
we have identified three distinct stationary
solutions, which will denote hereafter as A, B and C.
Solutions A and B exist at $\gamma_0=0$ and are characterized, at this limit,
in the first case by
a phase difference between the sites of $\pi$, so that $u(t)=w(t)\neq v(t)$; in the second case,
$v(t)=0$ and there is a phase difference of $\pi$ between the first and third
node, i.e. $u(t)=-w(t)$.
The third branch of solutions, namely C, exists for $\gamma_0\geq\sqrt{2\kappa^2 I_0^2(\gamma_1)-E^2}\approx1.4852$.
Interestingly, there is a qualitative difference in the bifurcation
diagram (an asymmetric pitchfork, which leads to an isolated branch and
a saddle-node bifurcation)
between the case examined in Refs.~\cite{pgk,kipnew} and the one considered
herein. In the former, the solutions A and B terminate through a saddle-node
and the C solution is the non-bifurcating branch of the broken pitchfork.
However, in our present system, it is the B and C solutions which cease
to exist
at the fold point of $\gamma_0=1.5741$,
whereas the non-bifurcating branch is now the A one. These results are
predicted by the averaged model and corroborated by the numerical analysis
of the modulated system and the corresponding numerically exact (up to
the prescribed tolerance of $10^{-12}$) time-periodic solutions.
Nevertheless, we have checked that for different values
of $\kappa'$ and $g'$, various features of the bifurcation diagram
may change. These include the above mentioned possibility of
A and B colliding rather than B and C, as well as even the possibility
of a fourth (D) branch of solutions emerging in the nonlinear system. The latter case is non-generic, and the bifurcation scheme strongly depends on $\kappa'$ and $g'$; for instance, at $\kappa'=0.1$ and $g'=I_0(2\gamma_1)$ the four branches exist at the Hamiltonian limit (with C and D branches corresponding to in phase solutions) and bifurcate branch A with C and B with D through saddle-nodes when $\gamma_0$ is increased.

\begin{figure}
\begin{center}
\begin{tabular}{cc}
    \multicolumn{2}{c}{\includegraphics[scale=0.4]{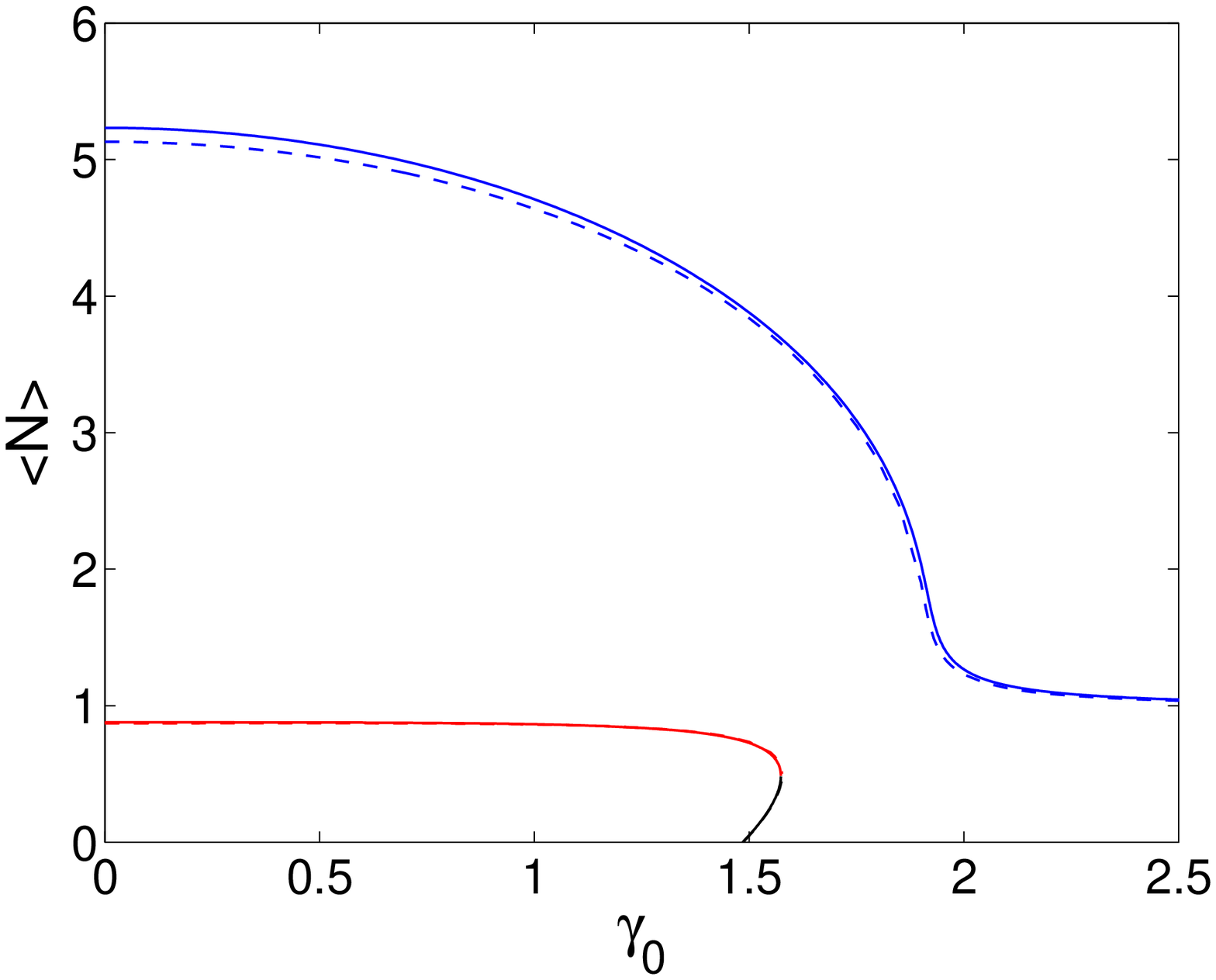}}\\
    \includegraphics[scale=0.4]{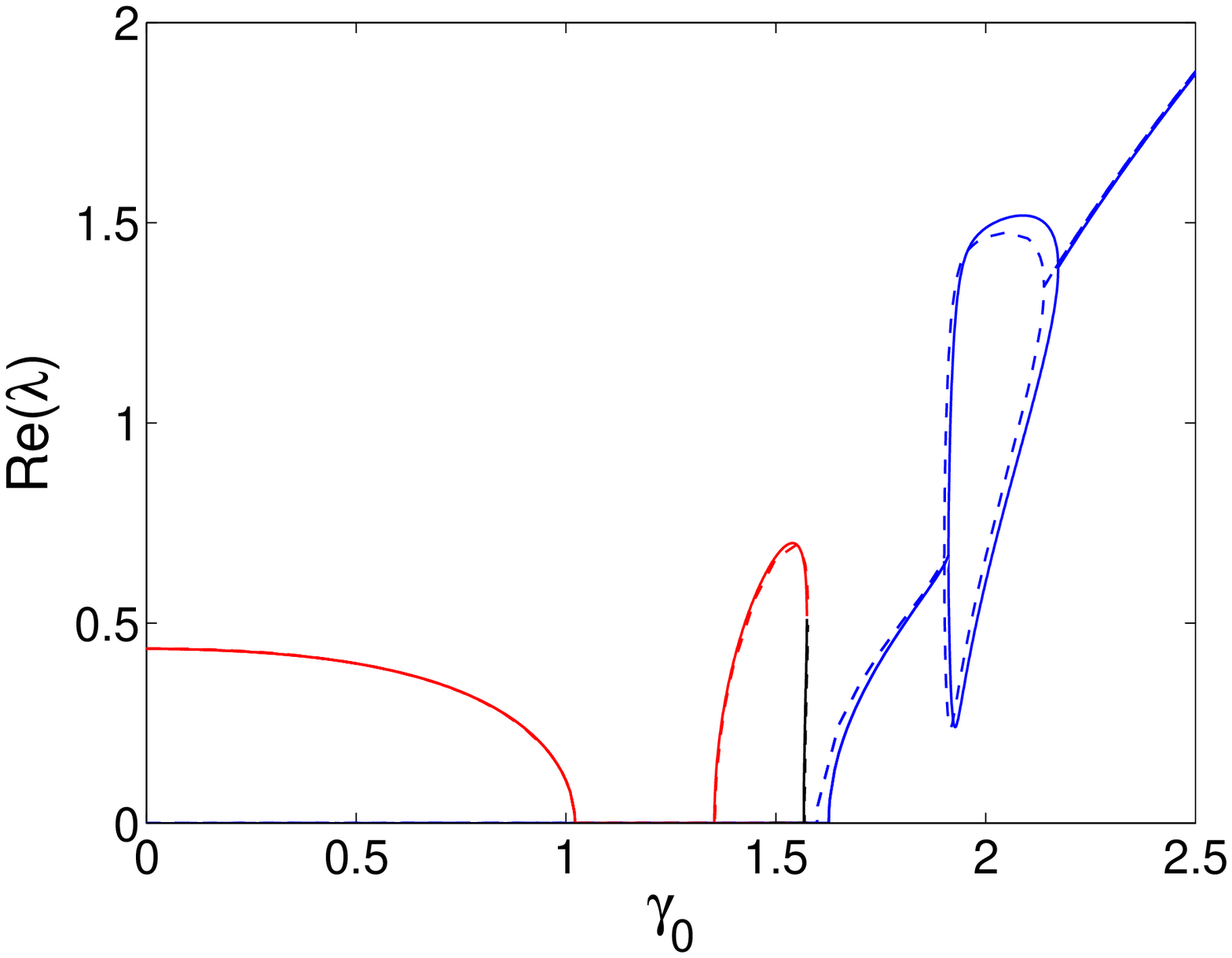} &
    \includegraphics[scale=0.4]{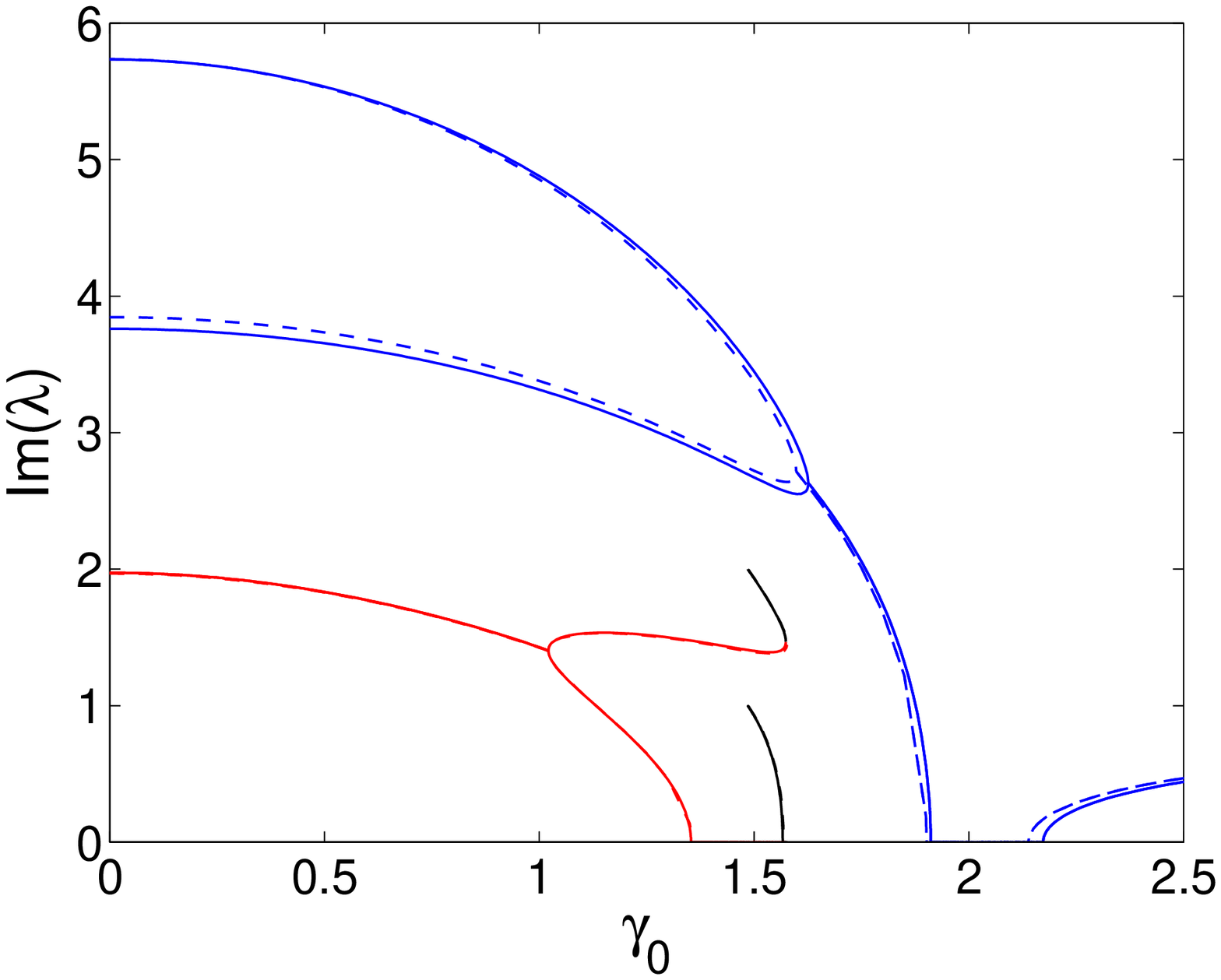} \\
\end{tabular}
\caption{(Color online) Averaged norm (top panel) and stability eigenvalues
(bottom panels) for a trimer with $\kappa=1$, $E=1$ and $\omega=\gamma_1=20$.
Solid (dashed) lines represent the values for the averaged
(non-autonomous) system,
whereas blue (red) line corresponds to the A (B) solution. The C solution is
depicted as a black line.} \label{fig:trimer}
\end{center}
\end{figure}

Figure~\ref{fig:trimer} shows both the averaged norm and the stability eigenvalues (imaginary and real parts),
together with the predicted values by the averaged equations.
Obviously, the prediction of the averaged system is excellent for the B and C solutions,
with a small discrepancy arising only for the A solution. We observe that the A solution
is stable for small $\gamma_0$, becoming unstable through a Hamiltonian
Hopf bifurcation~\cite{footnote2}
at $\gamma_0=1.5983$
($1.6261$ in the averaged system). The imaginary part of that quadruplet
of eigenvalues becomes zero (i.e., the instability becomes exponential) in
the range $\gamma_0\in[1.9014,2.1391]$ ($[1.9116,2.1731]$ for the averaged
system); the instability becomes again oscillatory above this range. The B
solution is oscillatorily unstable for small $\gamma_0$, becoming stable
via inverse Hopf bifurcation at $\gamma_0=1.0216$ ($1.0214$ in the
averaged case). The solution becomes exponentially unstable for
$\gamma\geq1.3552$ ($1.3534$, respectively for the effective autonomous
equation),  finally colliding with the C solution
and disappearing in the relevant saddle-node bifurcation
at $\gamma_0=1.5764$ ($1.5741$ in the averaged system) as explained above.
The C solution, which does not exist for $\gamma_0<1.4869$ ($1.4852$ in
the averaged system), is stable up to $\gamma_0=1.5687$ ($1.5667$, respectively
for the non-autonomous case), where it experiences an exponential bifurcation.
It is clear from the above comparisons that there is an excellent agreement
between the predictions of the original system and its effective,
averaged description.

\begin{figure}
\begin{center}
\begin{tabular}{cc}
    \includegraphics[scale=0.4]{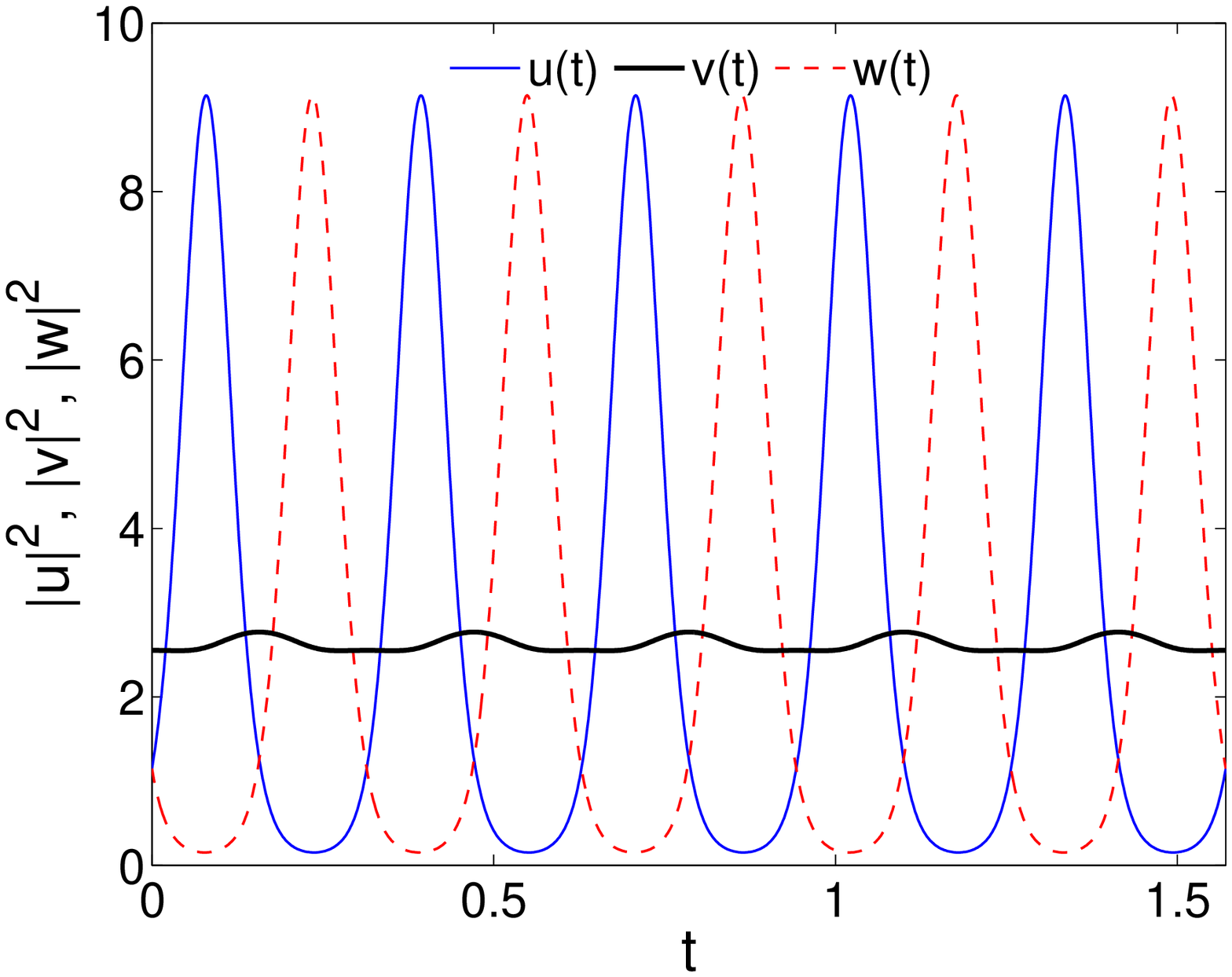} &
    \includegraphics[scale=0.4]{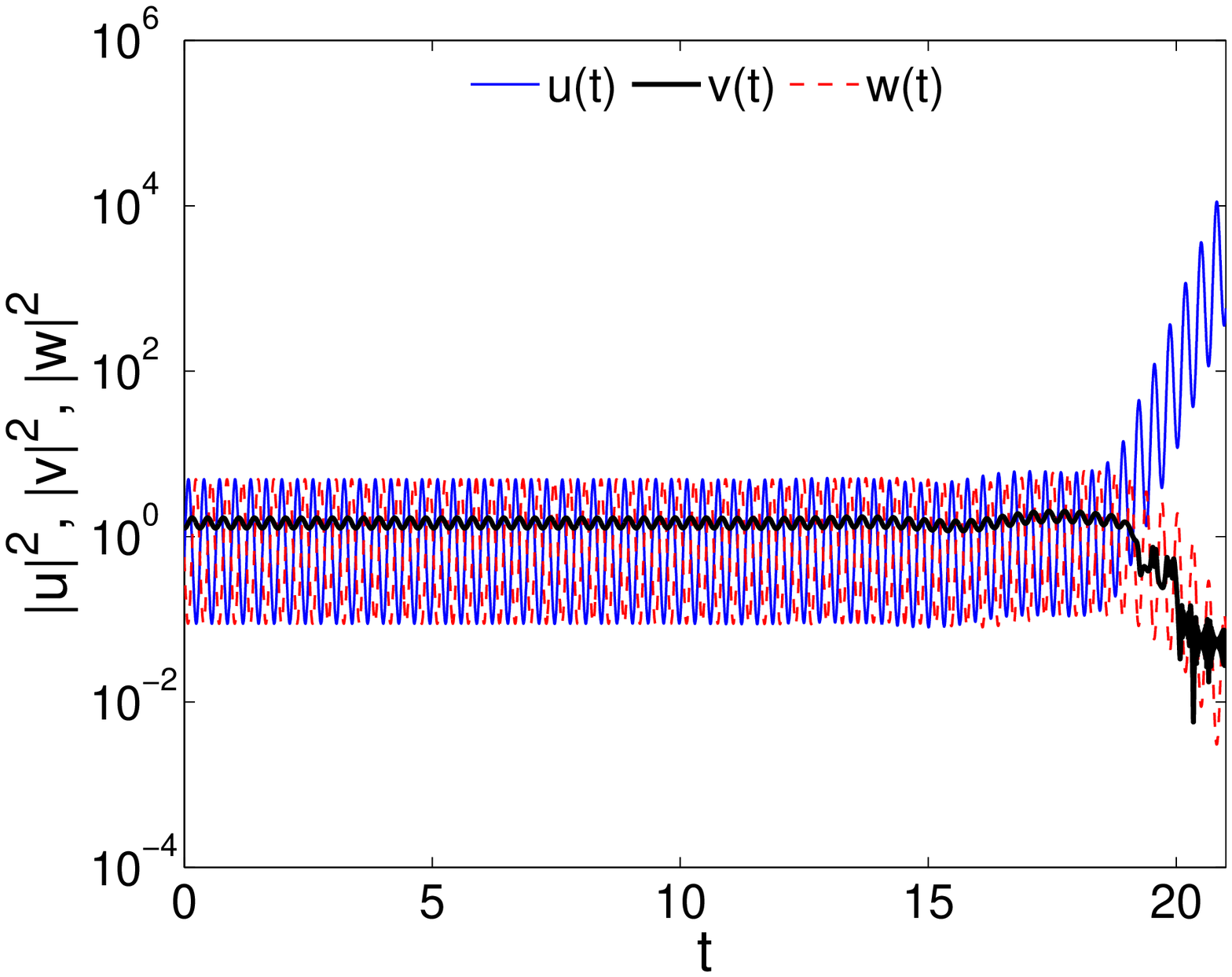} \\
    \includegraphics[scale=0.4]{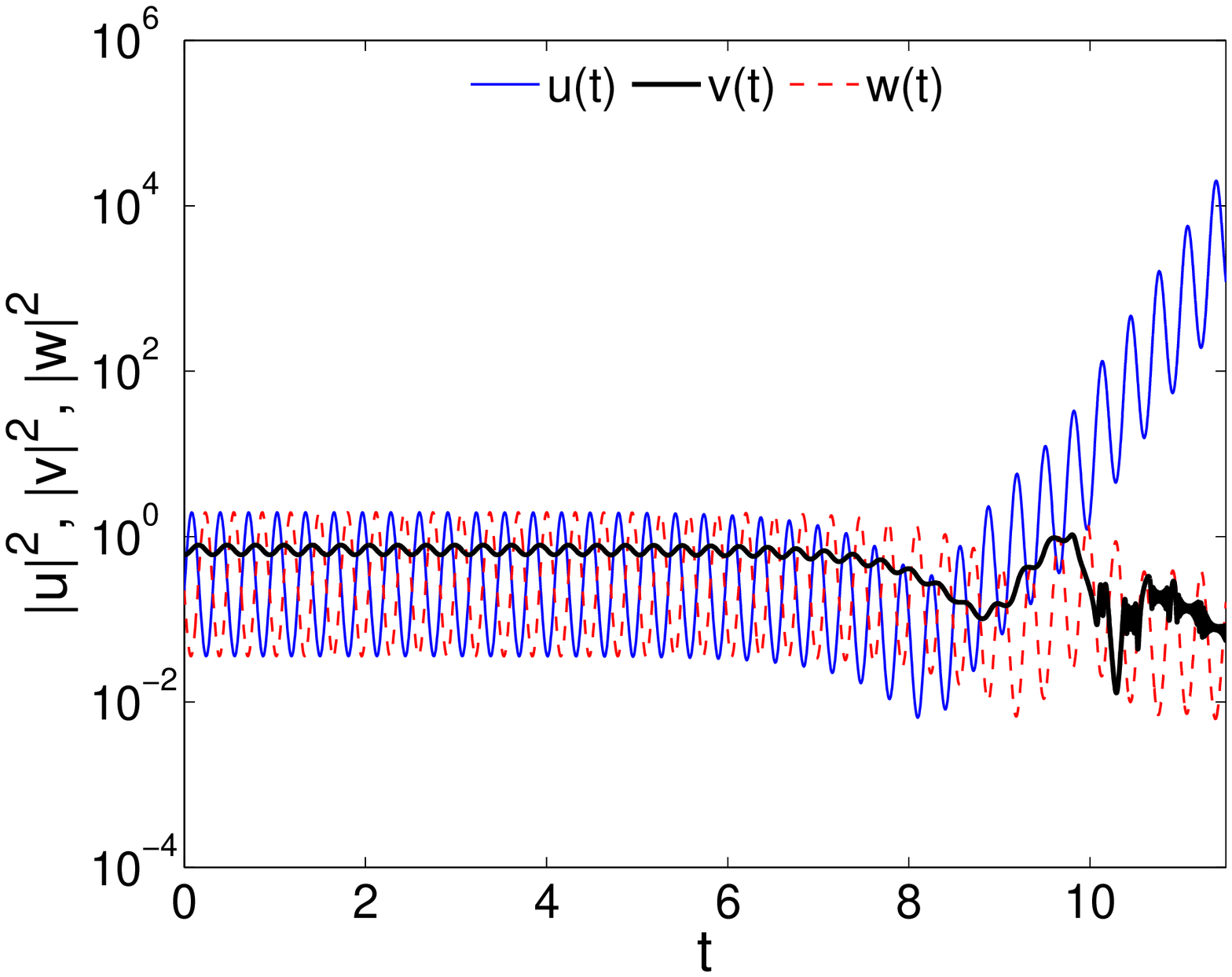} &
    \includegraphics[scale=0.4]{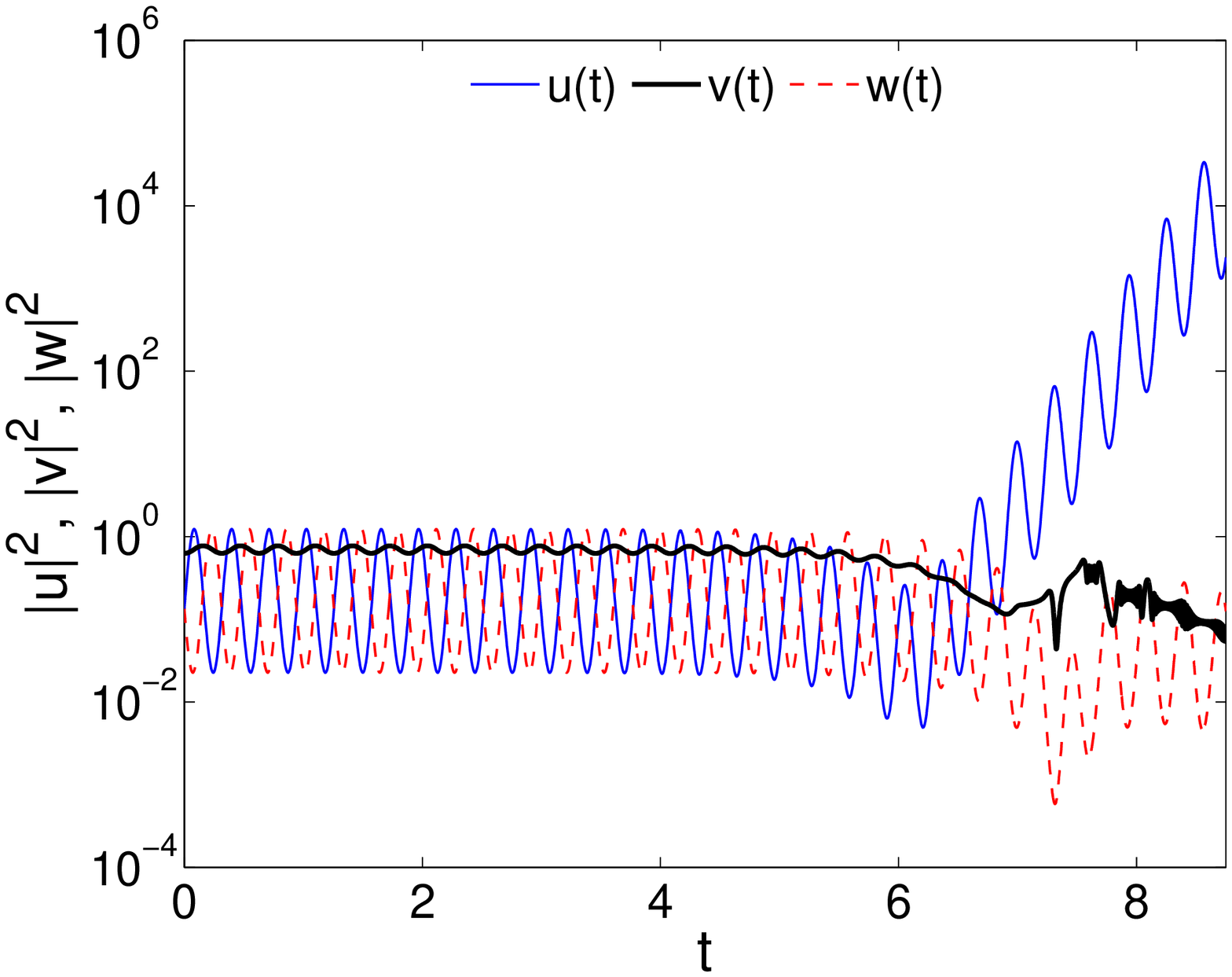} \\
\end{tabular}
\caption{(Color online) Dynamical evolution of A solutions for the modulated trimer in the case $\kappa=1$, $E=1$, $\gamma_1=1$ and $\omega=20$. The top left panel corresponds
to a stable evolution at $\gamma_0=0.5$; the top right panel shows the
evolution of an oscillatory unstable solution at $\gamma_0=1.8$;
the bottom left panel holds for an exponentially unstable solution at
$\gamma_0=2$; finally, the bottom right panel shows an oscillatorily unstable
solution at $\gamma_0=2.5$.} \label{fig:trimer_dynamics1}
\end{center}
\end{figure}

\begin{figure}
\begin{center}
\begin{tabular}{cc}
    \includegraphics[scale=0.4]{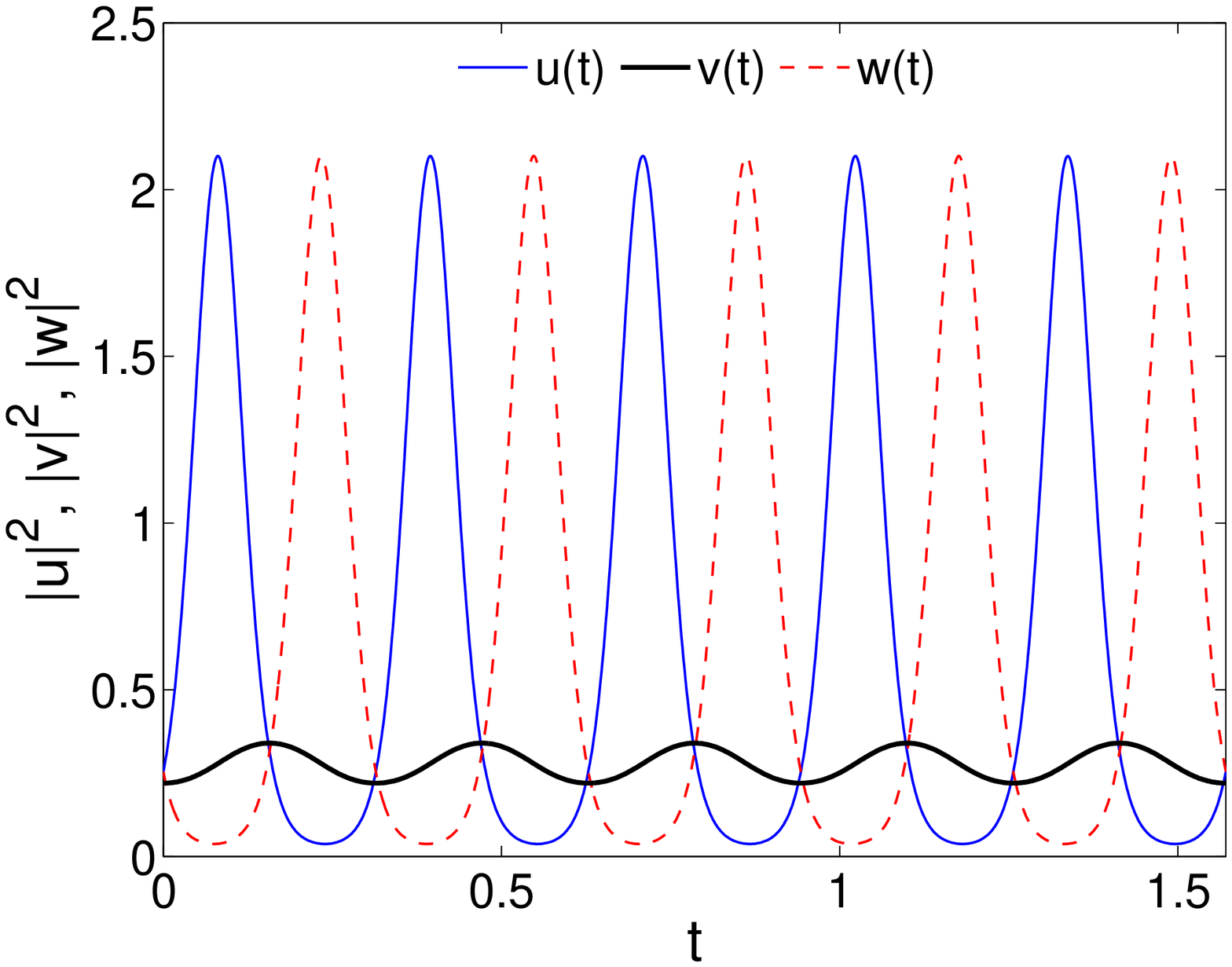} &
    \includegraphics[scale=0.4]{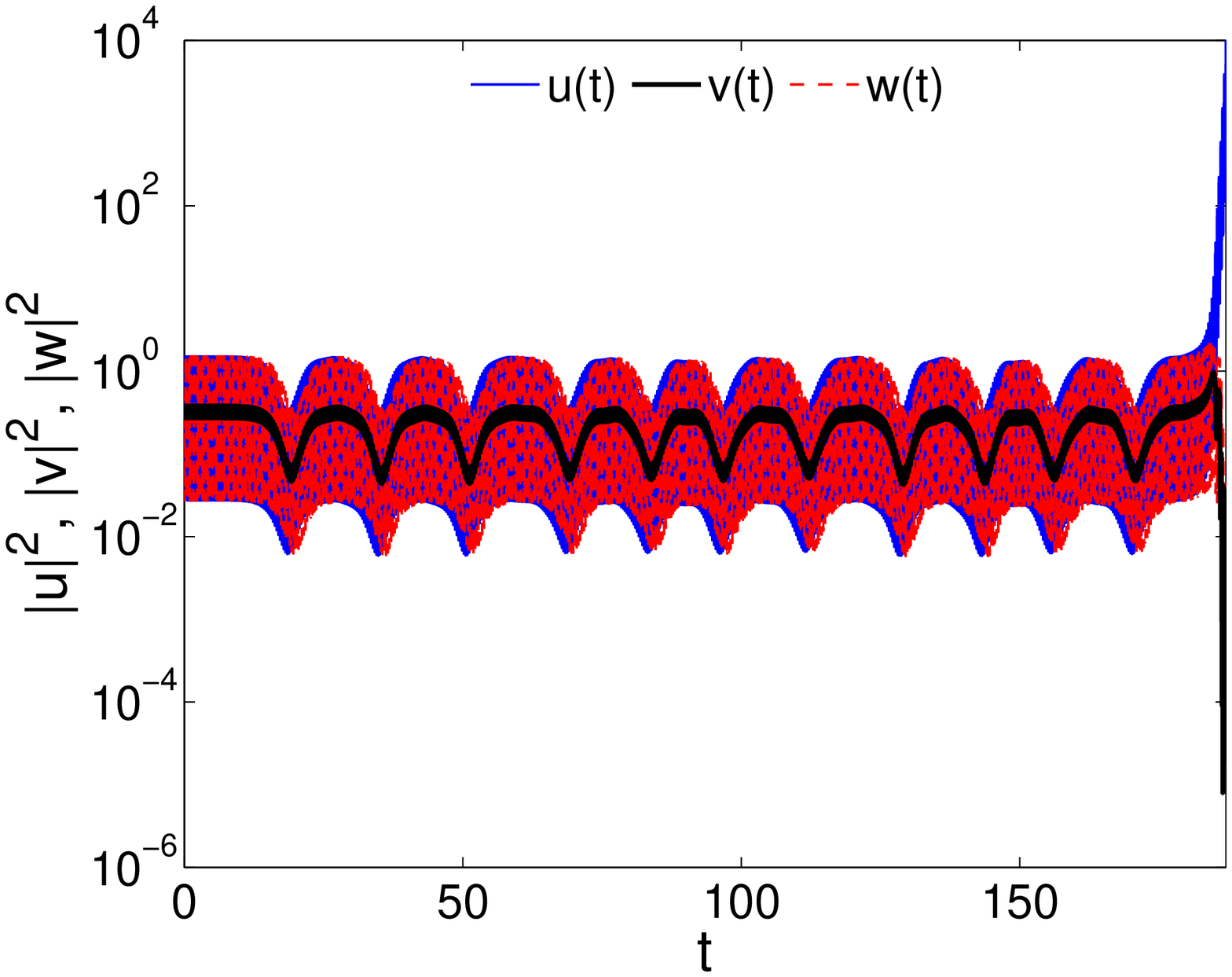} \\
    \includegraphics[scale=0.4]{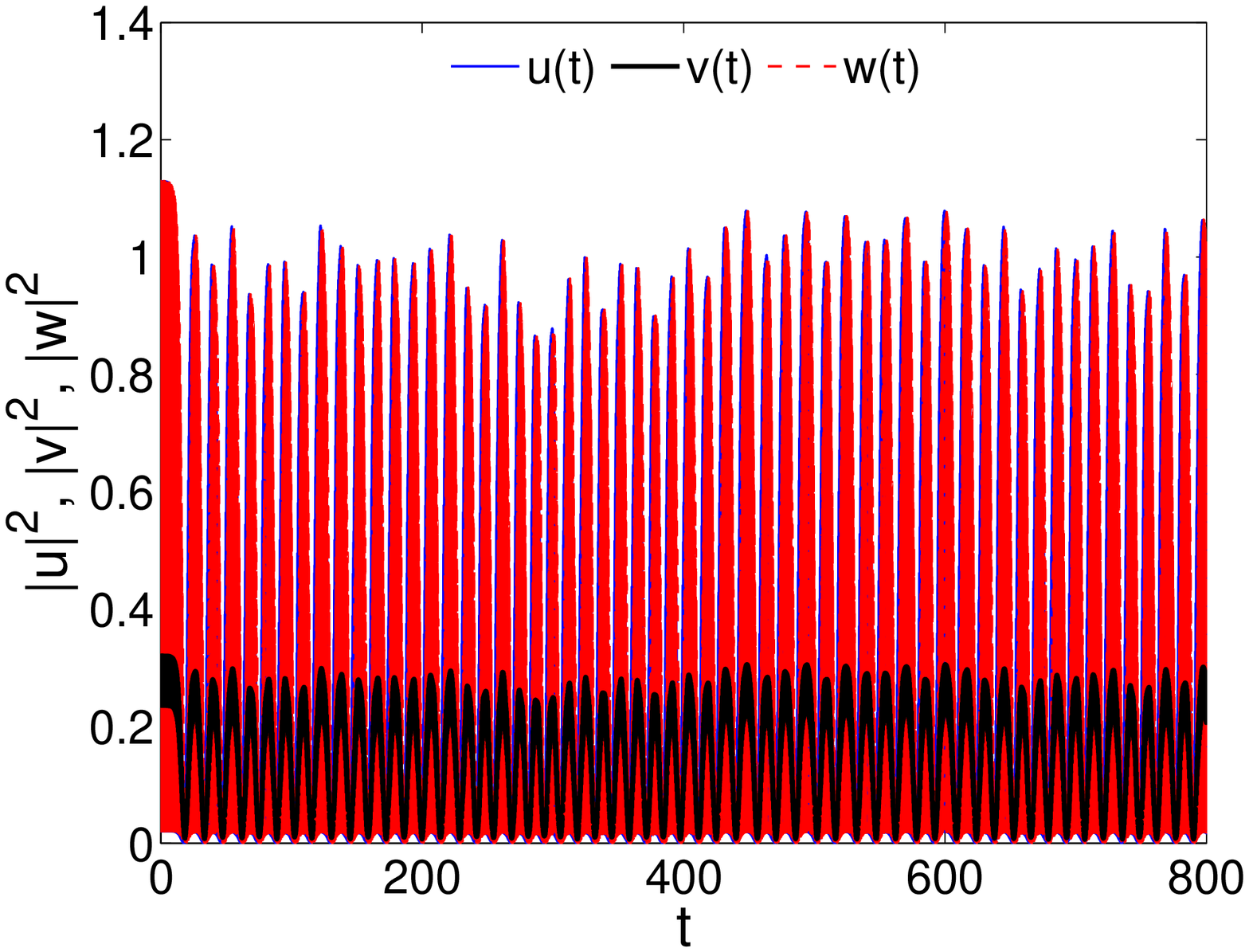} &
    \includegraphics[scale=0.4]{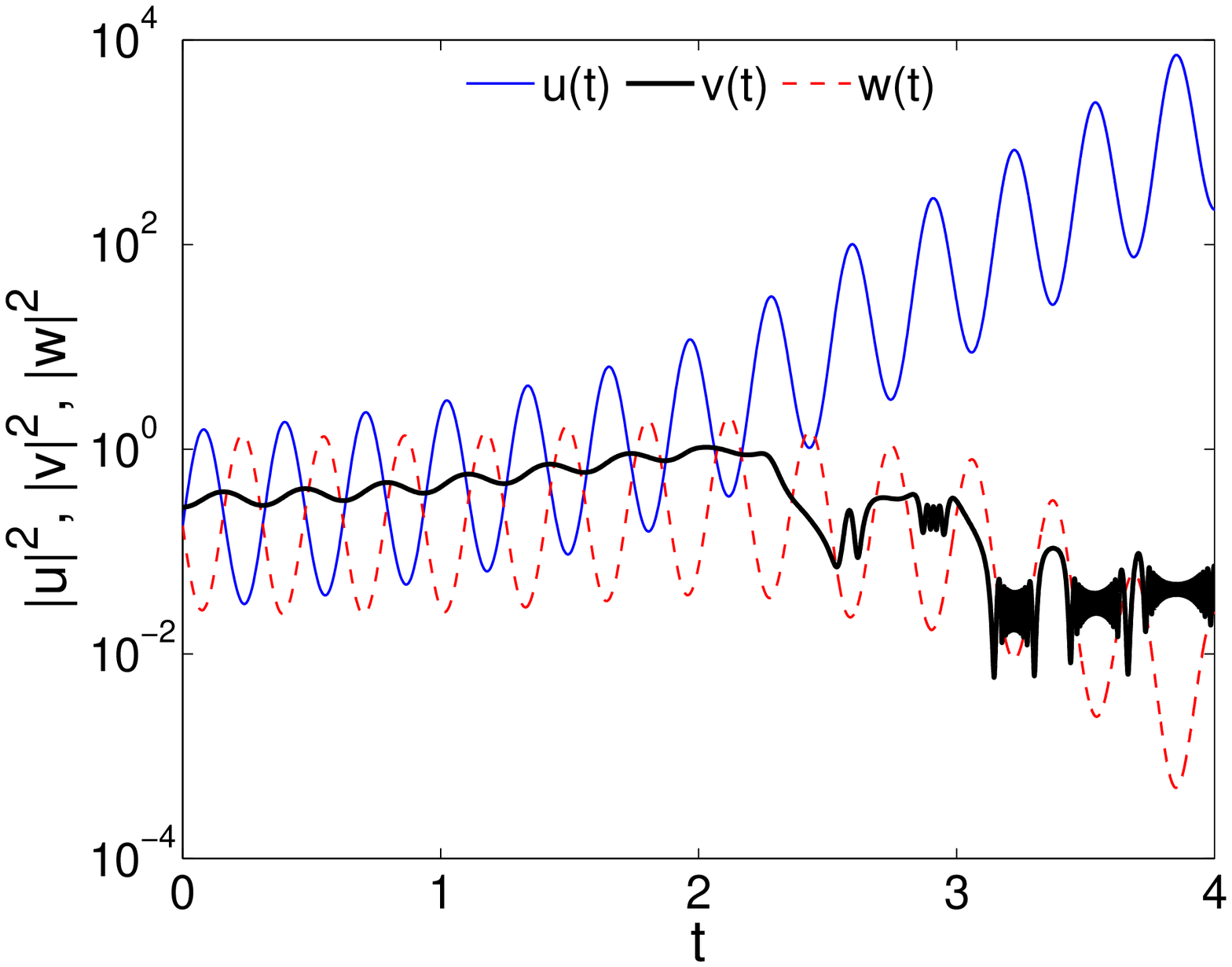} \\
\end{tabular}
\caption{(Color online) Dynamical evolution of B solutions for the
modulated trimer in the case $\kappa=1$, $E=1$, $\gamma_1=1$ and $\omega=20$.
The top left panel corresponds
to a stable evolution at $\gamma_0=1.2$; the top right panel shows the
evolution of an oscillatorily unstable solution at $\gamma_0=0.5$; the bottom
left  panel represents an exponentially unstable solution at $\gamma_0=1.57$;
finally, the bottom right panel shows the evolution at $\gamma_0=1.7$
(i.e. past the collision with branch C) using as initial
condition the solution at $\gamma_0=1.5$.}
\label{fig:trimer_dynamics2}
\end{center}
\end{figure}

\begin{figure}
\begin{center}
\begin{tabular}{cc}
    \includegraphics[scale=0.4]{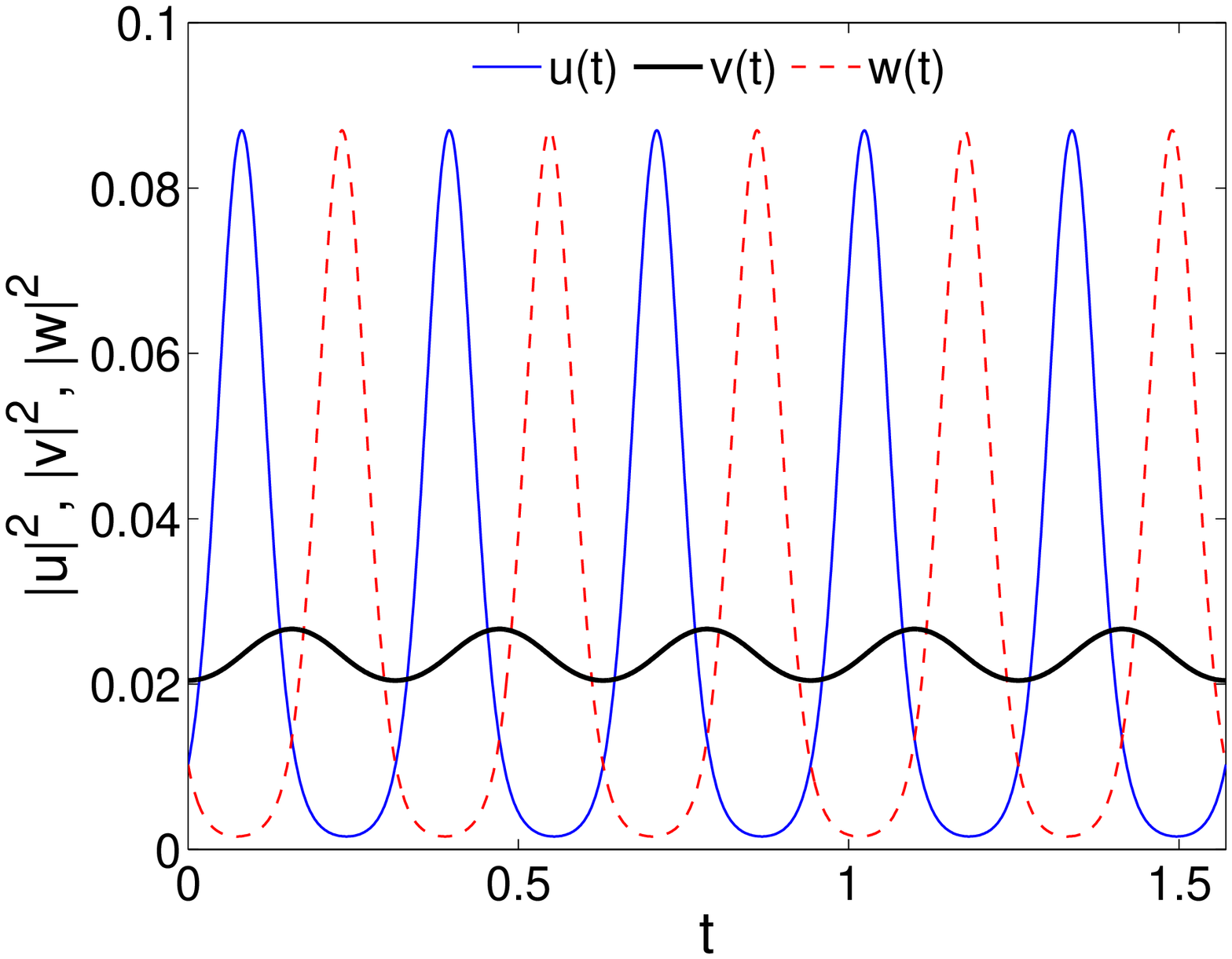} &
    \includegraphics[scale=0.4]{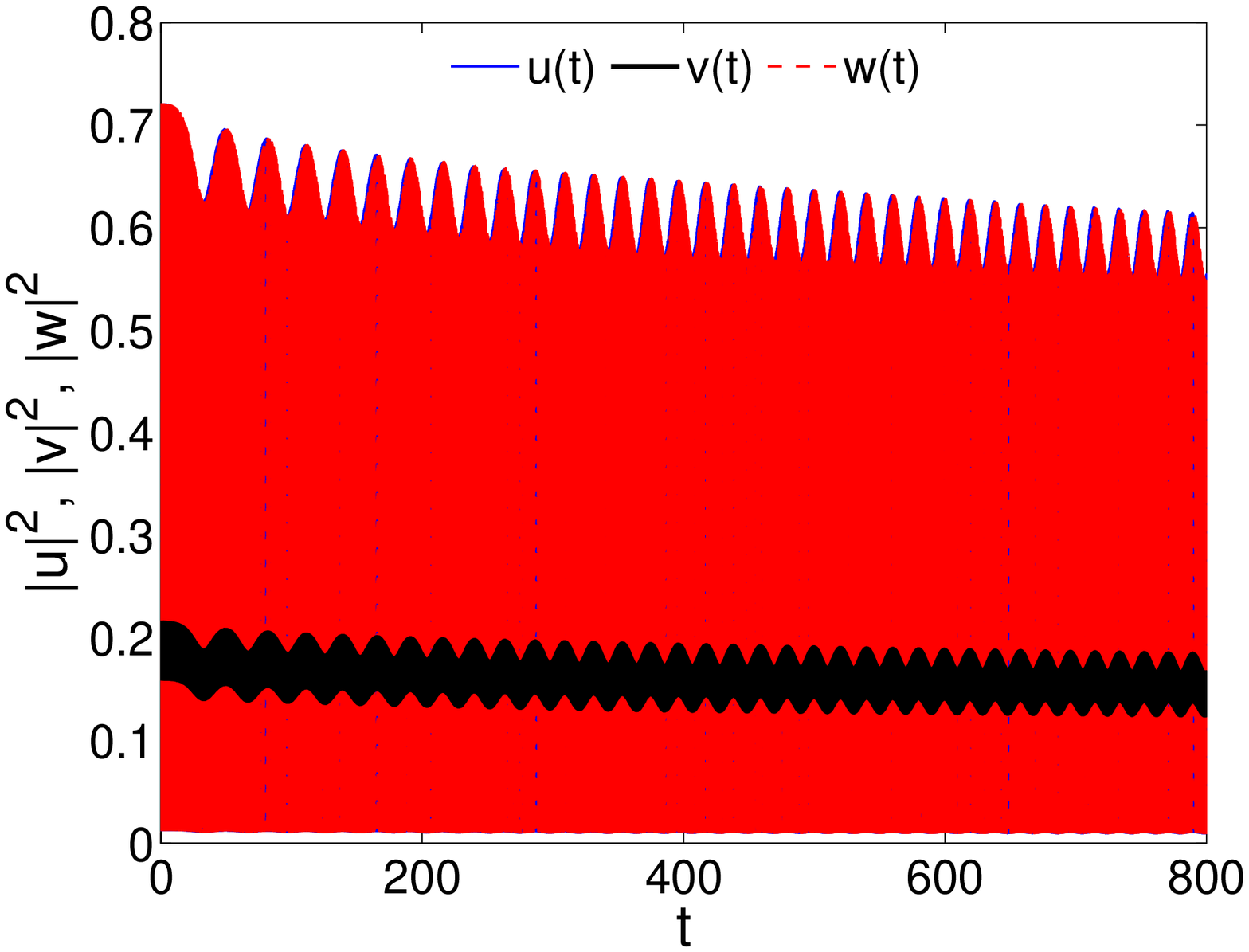} \\
    \includegraphics[scale=0.4]{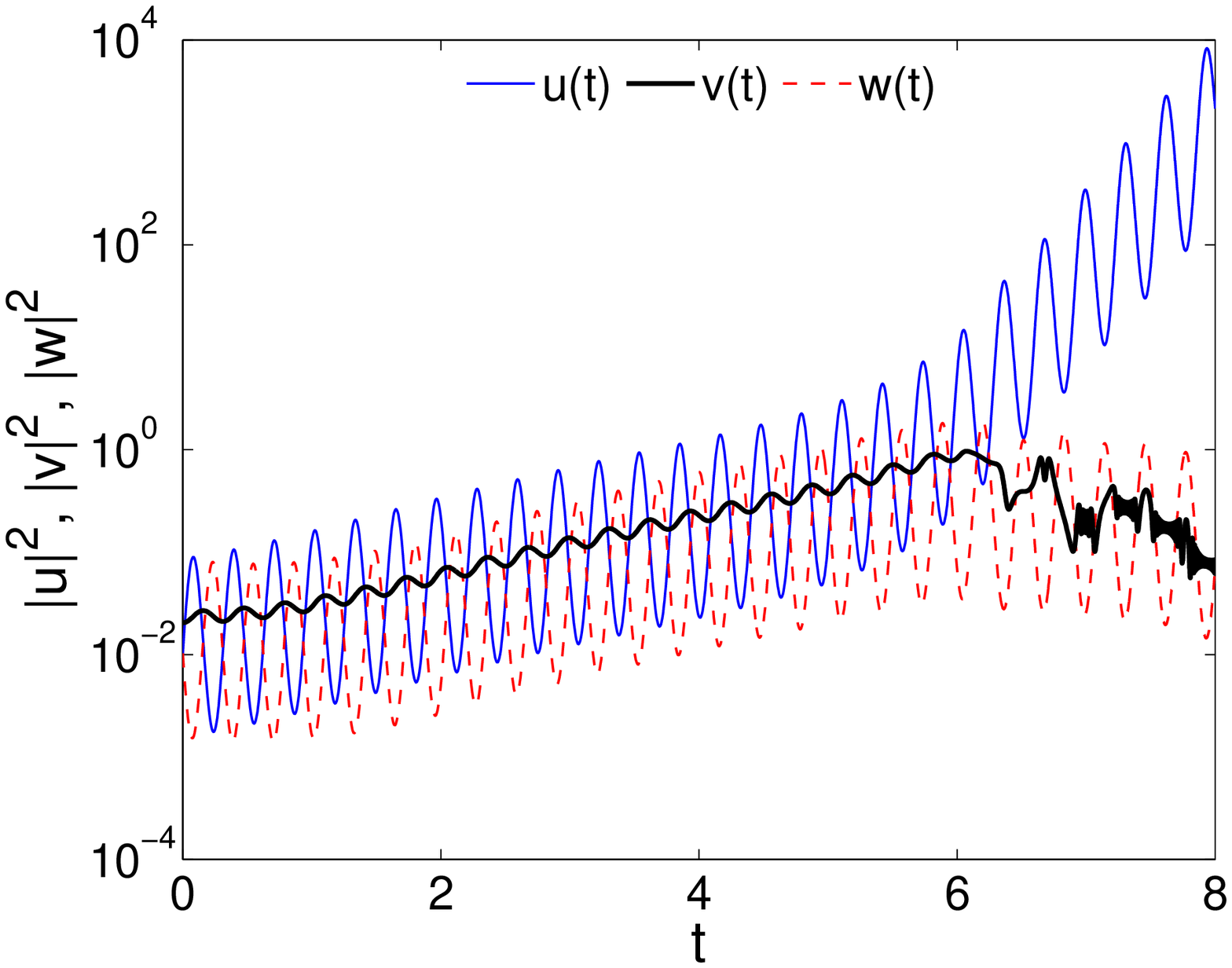} &
    \\
\end{tabular}
\caption{(Color online) Dynamical evolution of C solutions for the modulated trimer in the
case $\kappa_0=1$, $E=1$ $\gamma_1=1$ and $\omega=20$. The
top left panel corresponds to a stable evolution at $\gamma_0=1.5$; the top right panel shows the evolution of
an exponentially unstable solution at $\gamma_0=1.57$; the bottom left panel
corresponds to the case of $\gamma_0=1.7$ (i.e. past the saddle-node
bifurcation with branch B) using as initial condition the solution at
$\gamma_0=1.5$.} \label{fig:trimer_dynamics3}
\end{center}
\end{figure}

Figures~\ref{fig:trimer_dynamics1}-\ref{fig:trimer_dynamics3} show the dynamical evolution
of A, B and C solutions, respectively. In the case of
A solutions, we can observe their dynamical stability for
sufficiently low values of $\gamma_0$ (top left panel). As $\gamma_0$
is increased, initially an oscillatory (top right) and subsequently
an exponential (bottom left) instability arises. In the dynamical evolutions,
the fate of the solutions appears to be similar, with the gain site ultimately
growing, while the other two sites are eventually
observed to decay in amplitude.
Nevertheless, the exponential instability appears to manifest itself
faster, in consonance with our eigenvalue findings above.
As we progress to higher $\gamma_0$, an oscillatory instability arises
again, as shown in the bottom right panel but this time with a high
growth rate and a rapid destabilization accordingly.

The dynamics of the B branch is, arguably, somewhat more complex
in Fig.~\ref{fig:trimer_dynamics2}.
While a stable evolution for $\gamma_0=1.2$ is shown in the
top left panel, for smaller values of $\gamma_0$ (such as $\gamma_0=0.5$
of the top right panel), an oscillatory instability is present and appears
to lead to indefinite growth of the gain site, while the other two
sites decay in amplitude. Perhaps most intriguing is the case
of $\gamma_0=1.57$ of the bottom left panel of the figure. Here, the
dynamics does not appear to diverge, but rather seems to revert to a
quasi-periodic
motion,
yielding a bounded dynamical result. On the other hand, in the bottom right
case of $\gamma_0=1.7$, past the critical point of the
bifurcation with branch C, the dynamics is led to indefinite growth
(again with the gain site growing, while the other two are decaying
in amplitude).

Lastly, we consider different dynamical examples from within the
narrow interval of existence of branch C in Fig.~\ref{fig:trimer_dynamics3}.
In the top left panel case of $\gamma_0=1.5$, the stable evolution of
this branch is depicted. The exponential instability of the branch
in the top right panel for $\gamma_0=1.57$ appears, similarly
to branch B, to lead not to indefinite growth but rather to quasi-periodic
oscillation and a bounded dynamical evolution.
On the contrary, for values of
$\gamma_0$ past the saddle-node bifurcation with branch B (but similarly
to the dynamics of branch B for such values of $\gamma_0$), we observe
(cf. bottom left panel) indefinite growth in the dynamics for $\gamma_0=1.7$.

\section{Conclusions and Future Challenges}

In the present work, we have explored the potential of PT-symmetric
oligomer system (a dimer and a trimer, more concretely, although
generalizations to a higher number of sites are directly possible)
to have its gain/loss pattern periodically modulated in time.
Although this possibility may be somewhat more limited in optical
systems, it should in principle be possible in electric circuit settings.
As we argued, additionally, this kind of possibility may bear {\it significant}
advantages including most notably the {\it expansion} of the exact
PT-symmetric phase region. The latter threshold at the linear level
now becomes $\gamma_0 I_0(2 \gamma_1)$ or $\sqrt{2} \gamma_0 I_0(\gamma_1)$, for
the dimer and trimer, respectively, for a modulated gain/loss coefficient
with mean value
$\gamma_0$ and a periodic modulation of amplitude $\gamma_1$ and frequency $\omega$.
In addition to this expansion, we were able through our averaging
procedure to reduce
the non-autonomous full problem to an effective time-independent
(averaged) one, for which a lot of information (especially so
for the dimer case) can be obtained analytically, including the
existence and stability of the relevant solutions. The results
of the averaged equation approximation were generally found, in
the appropriate regime, to be in {\it excellent} agreement with
those of the full, time-dependent problem and its periodic solutions
and their Floquet exponents. This was the case both for the
(former) symmetric and asymmetric branches of the dimer and
the collision leading to their termination, but also for the branches
identified in the trimer, representing an apparent example
of an asymmetric pitchfork bifurcation.

One can envision
many interesting and relevant extensions
of the present work. One such would be to consider the
case where the PT-symmetry ``management'' could be applied
to a full lattice or to a chain of dimers, as in the work of~\cite{dmitriev2}.
There, it would be quite relevant to explore the impact of the
modulation to the solitary waves and localized solutions
of the lattice. Another possibility is to consider generalizations
of this ``linear'' PT-symmetry management towards a nonlinear
variant thereof.
More specifically, in the spirit
of~\cite{miron,konorecent,konorecent2,meid}, a PT-symmetric gain/loss term
could be applied to the nonlinear part of the dimer/trimer or lattice.
Then, one can envision a generalization of the notion of nonlinearity
management and of the corresponding averaging (see e.g.~\cite{pelizha,fkha}),
in order to formulate novel effective lattice media as a result of
the averaging. Such possibilities are currently under investigation
and will be reported in future publications.

\acknowledgments P.G.K. gratefully acknowledges support from
the US National Science Foundation (grant DMS-0806762 and
CMMI-1000337),
the Alexander von Humboldt Foundation and the US
AFOSR under grant FA9550-12-1-0332. P.G.K., R.L.H. and N.W. gratefully
acknowledge a productive visit to the IMA at the University of Minnesota.
J.C. acknowledges financial support from the
MICINN project FIS2008-04848.
The work of D.J.F. was partially supported by the Special Account for
Research Grants of the University of Athens.




\begin{thebibliography}{99}

\bibitem{R1} C.M. Bender and S. Boettcher,
Phys. Rev. Lett. {\bf 80}, 5243 (1998); C.M. Bender, S. Boettcher
and P.N. Meisinger, J. Math. Phys. {\bf 40}, 2201 (1999).


\bibitem{Muga} A. Ruschhaupt, F. Delgado, and J.G. Muga, J. Phys. A: Math. Gen. {\bf 38} (2005) L171.



\bibitem{ziad}
Z.H. Musslimani, K.G. Makris, R. El-Ganainy,
and D.N. Christodoulides, Phys. Rev. Lett. {\bf 100}, 030402 (2008);
K.G. Makris, R. El-Ganainy, D.N. Christodoulides, and Z.H. Musslimani,
Phys. Rev. A {\bf 81}, 063807 (2010).

\bibitem{Ramezani} H. Ramezani, T. Kottos, R. El-Ganainy, and
D.N. Christodoulides,
Phys. Rev. A {\bf 82}, 043803 (2010).

\bibitem{Kuleshov} M. Kulishov and B. Kress,
Optics Express {\bf 20}, 29319 (2012).

\bibitem{dncnat}
C.E. R{\"u}ter, K.G. Makris, R. El-Ganainy, D.N. Christodoulides, M. Segev, and D. Kip,
Nature Physics {\bf 6} (2010) 192.

\bibitem{salamo} A. Guo, G. J. Salamo, D. Duchesne, R. Morandotti,
M. Volatier-Ravat, V. Aimez, G. A. Siviloglou and D. N. Christodoulides,
Phys. Rev. Lett. {\bf 103}, 093902 (2009).

\bibitem{tsampikos_recent} J. Schindler,
A. Li, M.C. Zheng, F.M. Ellis, and T. Kottos,
Phys. Rev. A {\bf 84}, 040101 (2011).

\bibitem{tsampikos_review}
J. Schindler, Z. Lin, J. M. Lee, Hamidreza Ramezani, F. M. Ellis,
T. Kottos,
J. Phys. A: Math. Theor. {\bf 45}, 444029 (2012).


\bibitem{kot1} H. Ramezani, T. Kottos, R. El-Ganainy and D.N.
Christodoulides, Phys. Rev. A {\bf 82}, 043803 (2010).

\bibitem{sukh1} A.A. Sukhorukov, Z. Xu and Yu.S. Kivshar,
Phys. Rev. A {\bf 82}, 043818 (2010).

\bibitem{kot2} M.C. Zheng, D.N. Christodoulides, R. Fleischmann,
and T. Kottos, Phys. Rev. A {\bf 82}, 010103(R) (2010).

\bibitem{grae1} E.M. Graefe, H.J. Korsch, and A.E. Niederle,
Phys. Rev. Lett. {\bf 101}, 150408 (2008).

\bibitem{grae2} E.M. Graefe, H.J. Korsch, and A.E. Niederle,
Phys. Rev. A {\bf 82}, 013629 (2010).

\bibitem{kot3} Z. Lin, H. Ramezani, T. Eichelkraut, T. Kottos,
H. Cao, and D.N. Christodoulides, Phys. Rev. Lett. {\bf 106}, 213901 (2011).

\bibitem{pgk} K. Li and P. G. Kevrekidis
Phys. Rev. E {\bf 83}, 066608 (2011).

\bibitem{dmitriev1} S.V. Dmitriev, S.V. Suchkov, A.A. Sukhorukov,
and Yu.S. Kivshar,
Phys. Rev. A {\bf 84}, 013833 (2011).

\bibitem{dmitriev2} S.V. Suchkov,  B.A. Malomed, S.V. Dmitriev and
Yu.S. Kivshar, Phys. Rev. E {\bf 84}, 046609 (2011).

\bibitem{R30add1} R. Driben and B. A. Malomed, Opt. Lett. {\bf 36}, 4323 (2011).

\bibitem{R30add2} R. Driben and B. A. Malomed, Europhys. Lett. {\bf 96}, 51001
(2011).

\bibitem{R30add3} F. Kh. Abdullaev, V.V. Konotop, M. \"Ogren and M. P.
S{\o}rensen, Opt. Lett. {\bf 36}, 4566 (2011).

\bibitem{R30add4} N.V. Alexeeva, I.V. Barashenkov, A.A. Sukhorukov, and
Yu.S. Kivshar, Phys. Rev. A {\bf 85}, 063837 (2012).

\bibitem{R30add5} A.A. Sukhorukov, S.V. Dmitriev and Yu.S. Kivshar,
Opt. Lett. {\bf 37}, 2148 (2012).

\bibitem{R34} H. Cartarius and G. Wunner, Phys. Rev. A {\bf 86}, 013612 (2012);
J. Phys. A: Math. Theor. {\bf 45}, 444008 (2012).

\bibitem{R44} E.-M. Graefe, J. Phys. A: Math. Theor. {\bf 45}, 444015 (2012).


\bibitem{R46} A.S. Rodrigues, K. Li, V. Achilleos, P.G. Kevrekidis,
D.J. Frantzeskakis, and C.M. Bender, Rom. Rep. Phys. {\bf 65}, 5 (2013).

\bibitem{baras1} I. V. Barashenkov, S.V. Suchkov, A.A. Sukhorukov,
S.V. Dmitriev, and Yu.S. Kivshar,
Phys. Rev. A {\bf 86}, 053809 (2012).

\bibitem{baras2} I.V. Barashenkov, L. Baker, and N.V. Alexeeva
Phys. Rev. A {\bf 87}, 033819 (2013).

\bibitem{konorecent3} D.A. Zezyulin and V.V. Konotop,
Phys. Rev. Lett. {\bf 108}, 213906 (2012).


\bibitem{leykam} D. Leykam, V.V. Konotop, and A.S. Desyatnikov,
Opt. Lett. {\bf 38}, 371 (2013).

\bibitem{konous} K. Li, D. A. Zezyulin, V. V. Konotop, and P. G. Kevrekidis
Phys. Rev. A {\bf 87}, 033812 (2013).

\bibitem{djf} V. Achilleos, P. G. Kevrekidis, D. J. Frantzeskakis, and
R. Carretero-Gonz{\'a}lez, Phys. Rev. A {\bf 86}, 013808 (2012);
V. Achilleos, P.G. Kevrekidis, D.J. Frantzeskakis,
R. Carretero-Gonz{\'a}lez, arXiv:1202.1310.


\bibitem{kondark} Yu.V. Bludov, V.V. Konotop, and B.A. Malomed,
Phys. Rev. A {\bf 87}, 013816 (2013).

\bibitem{uwe} K. Li, P.G. Kevrekidis, B.A. Malomed, and U. G{\"u}nther,
J. Phys. A: Math. Theor. {\bf 44}, 444021 (2012).

\bibitem{tyugin} P.G. Kevrekidis, D.E. Pelinovsky, and D.Y. Tyugin,
arXiv:1303.3298, SIAM J. Appl. Dyn. Sys. (in press, 2013);
P.G. Kevrekidis, D.E. Pelinovsky, and D.Y. Tyugin, J. Phys. A: Math. Theor. {\bf 46}, 365201 (2013).
arXiv:1307.2973.

\bibitem{pickton} J. Pickton and H. Susanto, arXiv:1307.2788.

\bibitem{miron} A.E. Miroshnichenko, B.A. Malomed, and Yu.S. Kivshar
Phys. Rev. A {\bf 84}, 012123 (2011).

\bibitem{konorecent} F.Kh. Abdullaev, Y.V. Kartashov, V.V. Konotop,
and D.A. Zezyulin, Phys. Rev. A {\bf 83}, 041805 (2011).


\bibitem{konorecent2} D. A. Zezyulin, Y. V. Kartashov, V. V. Konotop,
Europhys. Lett. {\bf 96}, 64003 (2011).

\bibitem{meid} M. Duanmu, K. Li, R.L. Horne, P.G. Kevrekidis and N.
Whitaker, Phil. Trans. R. Soc. A {\bf 371}, 20120171 (2013).

\bibitem{boris_book} B.A. Malomed, {\it Soliton Management in Periodic
Systems} (Springer-Verlag, Berlin, 2006).

\bibitem{turitsyn_review} S.K. Turitsyn, B.G. Bale, and M.P. Fedoruk,
Phys. Rep. {\bf 521}, 135 (2012).
%

\bibitem{footnote} Although it is entirely
straightforward to envision a modulation of the nonlinear potential too,
but this will be deferred as a separate topic for future study.


\bibitem{kipnew} K. Li, P.G. Kevrekidis, D.J. Frantzeskakis,
C.E. R{\"u}ter, D. Kip, arXiv:1306.2255.

\bibitem{shooting} S. Flach and A. V. Gorbach, Phys. Rep. {\bf 467}, 1 (2008);
F. Palmero, L.Q. English, J, Cuevas, R. Carretero-Gonz\'alez, and P.G. Kevrekidis,
Phys. Rev. E {\bf 84}, 026605 (2011);
T. Cretegny and S. Aubry, Phys. Rev. B {\bf 55} R11929 (1997);
J. G\'omez-Garde\~{n}es, L.M. Flor\'{\i}a, M. Peyrard, and A.R. Bishop, Chaos {\bf 14}, 1130 (2004);
T.R.O. Melvin, A.R. Champneys, P.G. Kevrekidis and J. Cuevas, Physica D {\bf 237}, 551–567 (2008).




\bibitem{ptkg} J. Cuevas, P.G. Kevrekidis, A. Saxena and A. Khare, ArXiv: 1307.6047.


\bibitem{footnote2} It is interesting to note here that despite the fact
that the system is no longer Hamiltonian, the bifurcation
arising has all the characteristics of Hamiltonian Hopf
including the collision of two eigenvalues and the formation of
a quartet. We note that it would be especially interesting to consider
the question of the potential existence of an analogous notion to the
Krein signature
[see, e.g., R. S. MacKay, in \textit{Hamiltonian Dynamical Systems}, edited by R. S. MacKay
and J. Meiss (Hilger, Bristol, 1987), p.137]
for these PT-symmetric systems that
could predict such potential Hopf bifurcations and quartet formations.



\bibitem{pelizha}  D.E. Pelinovsky, P.G. Kevrekidis and D.J. Frantzeskakis,
Phys. Rev. Lett. {\bf 91}, 240201 (2003);
D.E. Pelinovsky, P.G. Kevrekidis, D.J. Frantzeskakis
and V. Zharnitsky, Phys. Rev. E {\bf 70}, 047604 (2004);
V. Zharnitsky and D.E. Pelinovsky, Chaos {\bf 15}, 037105 (2005);
P.G. Kevrekidis, D.E. Pelinovsky and A. Stefanov, J. Phys. A
{\bf 39}, 479 (2006).

\bibitem{fkha} F.Kh. Abdullaev, E.N. Tsoy, and B.A. Malomed, Phys. Rev. A {\bf 68},053606 (2013);
F.Kh. Abdullaev, P.G. Kevrekidis, and M. Salerno, Phys. Rev. Lett. {\bf 105}, 113901 (2010).



\end{thebibliography}
\end{document}